\documentclass[useAMS,usenatbib]{mn2e}
\usepackage{color}
\usepackage{graphicx}
\usepackage{lscape}
\usepackage{amsmath}

\newcommand\aj{\rmfamily{AJ}}
\newcommand\apj{\rmfamily{ApJ}}
\newcommand\apjs{\rmfamily{ApJS}}
\newcommand\aaps{\rmfamily{A\&AS}}
\newcommand\aap{\rmfamily{A\&A}}
\newcommand\mnras{\rmfamily{MNRAS}}

\title[Testing spectral models for stellar populations]{Testing spectral models for stellar populations with star
clusters: I. Methodology}

\author[Roberto Cid Fernandes and Rosa M. Gonz\'alez Delgado]{Roberto Cid Fernandes$^{1}$\thanks{E-mail:cid@astro.ufsc.br} and
Rosa M. Gonz\'alez Delgado$^{2}$\thanks{E-mail:
rosa@iaa.es}\\
$^1$Departamento de Fisica-CFM, Universidade Federal de Santa Catarina,
P.O. Box 476, 88040-900, Florian\'opolis, SC, Brazil\\
$^2$Instituto de Astrof\'{\i}sica de Andaluc\'{\i}a (CSIC), P.O. Box
3004, 18080 Granada, Spain}

\begin{document}

\date{2009 July}


\maketitle

\label{firstpage}

\begin{abstract}
High resolution spectral models for simple stellar populations (SSP)
developed in the past few years have become a standard ingredient in
studies of stellar population of galaxies. As more such models become
available, it becomes increasingly important to test them.  In this
and a companion paper, we test a suite of publicly available
evolutionary synthesis models using integrated optical spectra in the
blue-near-UV range of 27 well studied star clusters from the work of
Leonardi \& Rose (2003) spanning a wide range of ages and
metallicities.  Most (23) of the clusters are from the Magellanic
clouds.  This paper concentrates on methodological aspects of spectral
fitting. The data are fitted with SSP spectral models from Vazdekis
and collaborators, based on the MILES library. Best-fit and Bayesian
estimates of age, metallicity and extinction are presented, and
degeneracies between these parameters are mapped. We find that these
models can match the observed spectra very well in most cases, with
small formal uncertainties in $t$, $Z$ and $A_V$. In some cases, the
spectral fits indicate that the models lack a blue old
population, probably associated with the horizontal branch.  This
methodology, which is mostly based on the publicly available code {\sc
starlight}, is extended to other sets of models in Paper II, where a
comparison with properties derived from spatially resolved data
(color-magnitude diagrams) is presented. The global aim of these two
papers is to provide guidance to users of evolutionary synthesis
models and empirical feedback to model makers.
\end{abstract}


\section{Introduction}
\label{sec:Introduction}

Star clusters (SCs) are ideal test beds for the evolution of stars and
stellar populations. Their key properties are their age ($t$) and
metallicity ($Z$), but extinction ($A_V$) also plays a role in
determining observed properties.  The classical method to estimate
these properties is through the comparison of observed color-magnitude
diagrams (CMDs) to the predictions of stellar evolution models.  For
distant SCs, however, only integrated light measurements are
available, and the traditional approach in this case is to use broad
band colors or spectral indices to estimate their properties.  Both
these methods have a long history in the literature (see, e.g., the
conference books Lamers et al.\ 2004; P\'erez et al.\ 2009; Geisler et
al.\ 2001).

A third and in principle more thorough alternative has become possible
with the availability of high spectral resolution evolutionary
synthesis models, which have appeared in the literature in the past
few years (see e.g. Bruzual 2007; Gonz\'alez Delgado 2009 and
references therein). These models allow the fitting of observed
spectra on a $\lambda$-by-$\lambda$ basis, thus incorporating all
available information.  Curiously, though numerous studies have used
these models to disentangle the mixture of stellar populations in
galaxies (e.g., Cid Fernandes et al.\ 2007 and references therein),
relatively little has been done on the much simpler problem of
applying spectral synthesis to infer the properties of SCs.  The most
complete study in this vein has just been published by Koleva et al.\
(2008, K08), who analyzed optical spectra of Galactic globular
clusters from the Schiavon et al.\ (2005) atlas. They find that
spectral fitting provides $t$ and $Z$ estimates in very good agreement
with those obtained through a CMD analysis, thus validating spectral
synthesis as a useful tool for SC work.  Similar work, but focusing on
the near-IR range, has been recently published by Lan\c{c}on et al.\
(2009).

This paper is dedicated to methodological aspects of spectral fitting
of SCs. Detailed fits of optical spectra are used to estimate $t$, $Z$
and $A_V$ for 27 SCs from the sample of Leonardi \& Rose (2003;
hereafter LR03). A thorough mapping of the parameter space is
performed to quantify the uncertainties and degeneracies in the
parameters. All results presented here are based on fits obtained with
the Vazdekis-MILES single stellar populations (SSP) models.  In a
companion paper (Gonz\'alez Delgado \& Cid Fernandes 2009, Paper II)
the same methodology is extended to a suite of other publicly
available high resolution evolutionary synthesis models, allowing us
to evaluate their pros and cons in a comparative fashion, as well as
to quantify the uncertainties in SC properties resulting from variance
in the ingredients for the analysis.  Paper II also compares the SC
properties derived from spectral fitting to those obtained from CMD
work on the same clusters.

The central goal of these two papers is to provide useful
guidance to users of evolutionary synthesis models and spectral
synthesis tools, as well as empirical feedback to model makers. Given
these goals, we deliberately restrict our analysis to models and
analysis tools as found in public distributions, with as little as
possible manipulation. We also point out that although this study
focus entirely on SCs, its underlying motivation is to evaluate the
reliability of methods and ingredients widely disseminated in modern
analysis of the integrated light of galaxies (see Cid Fernandes 2007
and references therein). SCs are the undoubtedly the best test-beds
for this. If SC ages can be accurately recovered with current models
and spectral synthesis methods, this lends confidence to applications
which seek to estimate the star formation histories of
galaxies. Similarly, if metallicities are also well recovered, then
chemical evolution studies are warranted. In short, establishing the
accuracy with which age and metallicity of SCs can be derived through
spectral synthesis is important to understand the systematics and
limitations of studies of galaxy evolution based on the analysis of
the fossil record encoded in their spectra.

The SC data and SSP models are described in Sections \ref{sec:Data}
and \ref{sec:STPop_Models} respectively. Section
\ref{sec:SpecFitMethod} describes the spectral fitting method.
Appendices \ref{app:Appendix_ZIwork} and \ref{app:Appendix_SNwork}
describe technical aspects of the fits and simulations to evaluate the
effects of noise.  Results of these fits are presented in Section
\ref{sec:SpecFitResults}, which also illustrates and discusses the
degeneracies involved.  Section \ref{sec:Bayes} applies a Bayesian
formalism to provide formal estimates of the uncertainties in the
derived parameters.  Finally, Section \ref{sec:Summary} summarizes our
main results.

\section{Data}
\label{sec:Data}

Spectra for 23 SCs in the Magellanic Clouds (3 in the SMC and 20 in
LMC) and 4 Galactic clusters (GCs) were collected by LR03 with the
1.5m telescope at the Cerro Tololo Inter-American Observatory, and
kindly made available to us. The spectra cover the wavelength range
3500--4700 \AA\ with a final resolution FWHM$_{inst} = 5.7$ \AA\
(including the gaussian smoothing applied by LR03). Our analysis
restricts the spectral range to 3650--4600 \AA, an interval covered by
all available SSP models, and which skips calibration problems towards
the edges of some of the spectra.  The signal-to-noise ratio of
these spectra, measured from the rms deviation of $F_\lambda$ in the
4010--4060 \AA\ window, range from 19 to 112, with a sample average
$S/N$ of 54.

In addition to the high quality of the spectra, the motivation to
re-analyze these data with a spectral fitting approach stems from the
fact that their $t$'s and $Z$'s were determined through independent
means, including CMDs, spectral indices, colors and spectroscopy of
individual stars (LR03).  These independent estimates allow a
comparative analysis similar to that performed by K08. A further
motivation is that, compared to the systems studied by K08, these SCs
are generally younger (from less than 0.1 to 3 Gyr according to LR03),
which makes $Z$ estimates more challenging (Wolf et al.\ 2007).

\section{Evolutionary Synthesis Models}
\label{sec:STPop_Models}

Evolutionary tracks and stellar libraries are the two most important
ingredients in evolutionary synthesis models. As a result of recent
work in both fronts, there are nowadays several sets of models
available for spectral synthesis work.  One of our major goals is to
subject these models to an empirical test. This is done in Paper II,
where results obtained with different publicly available models are
compared. For clarity, this paper uses only one version of these
models.

We chose to work with the set of SSP models built by Vazdekis et al.\
(2009), available at
http://www.ucm.es/info/Astrof/miles/models/models.html.  The main
novelty in these models is that they incorporate the MILES library of
S\'anchez-Bl\'azquez et al.\ (2006).  MILES contains about one
thousand stars spanning a large range of stellar parameters. The
spectra cover the 3540 to 7410 \AA\ range with a resolution of 2.3
\AA\ (FWHM). The spectra are well calibrated in flux and
wavelength, and represent a significant improvement with respect to
previous libraries in the coverage of the metallicity, number of giant
stars and other aspects. (The ELODIE library of Le Borgne et al.\ 2004
is similar in size and coverage of stellar parameters, but, starting
at 4000 \AA, does not cover the range covered by our data.)

This library was incorporated into the evolutionary synthesis code of
Vazdekis (1999), using ``Padova 2000'' evolutionary tracks (Girardi et
al.\ 2000, 2002) and a Salpeter IMF (Salpeter 1955). SSP spectra were
computed for $N_Z = 6$ metallicities: $Z = 0.0004$, 0.001, 0.004,
0.008, 0.019 (solar) and 0.03, and $N_t = 46$ ages between 0.10 and
17.78 Gyr.  The lack of younger ages is due to the small number of hot
(over 15000 K) stars in MILES. Few SCs in our sample are young enough
to be affected by this limitation.

As with other sets of models (see Paper II) the sampling in age
is practically continuous, but predictions are only available for a
few metallicities.  Model-users are thus limited to a coarse sampling
in $Z$. Rigorously speaking, the fact that evolutionary stages occur
at different times for different $Z$'s makes interpolation to a finer
grid physically invalid, even though reasonable approximations can be
obtained.  Given our goal to remain as faithful as possible to
SSP-models in their original form, available to users in general, we
concentrate on results obtained with the original $N_Z = 6$ grid.
Experiments with $Z$-interpolated SSP spectra are reported in Appendix
\ref{app:Appendix_ZIwork}.

Throughout the rest of this paper ages will be given in yr, and
metallicities will be quoted in a log-solar scale.  In this notation,
the $Z$ values above correspond to (rounding up to the first decimal)
$\log Z/Z_\odot = -1.7$, -1.3, -0.7, -0.4, 0, and +0.2. For
consistency with the notation in Paper II, this set of models will be
referred to as ``V00s'' (an acronym for ``Vazdekis $+$ Padova 2000 $+$
Salpeter'').

\section{Spectral fitting: Method}
\label{sec:SpecFitMethod}

The goal of spectral fitting is to find a model spectrum ($M_\lambda$)
which best matches the observed one ($O_\lambda$), taking into account
all $\lambda$'s. The fits presented below were carried out with
version 05 of the publicly available {\sc
starlight}\footnote{www.starlight.ufsc.br} code, better known as a
tool to retrieve the star formation history of galaxies by fitting a
spectrum with a {\em mixture} of $N_\star$ SSPs from a base spanning
wide ranges of $t$ and $Z$ (see Cid Fernandes et al 2005, 2009; Asari
et al 2007 and the {\sc starlight} user manual for examples and
references).

A so far unused feature of this code is that it also fits $O_\lambda$
with each of the base spectra {\em individually}, thus making it
useful to study single-population systems like SCs. This section
describes the parameters and technical aspects of these fits. The
level of details is justified by the fact that this is the first time
{\sc starlight} is used to fit SCs.

A series of technical details lurk behind the deceiving
simplicity of fitting an observed SC spectrum.  Should kinematical
parameters be fitted when they are not relevant?  How many degrees of
freedom are involved in a fit? What to do when no measure of the error
in $O_\lambda$ is available?  How do noise affects the fits?  Should
one work with a discrete base of SSPs or a $\sim$ continuous one
obtained through interpolation?  Some of these issues are discussed
below, while others are addressed in Appendices
\ref{app:Appendix_ZIwork} and \ref{app:Appendix_SNwork}.  Because of
the general goals outlined in Section \ref{sec:Introduction}, our
default strategy to tackle these issues is to mimic inasmuch as
possible the way {\sc starlight} is used to analyze galaxy spectra,
while at the same time avoiding manipulations of the original SSP
models. Variations over this strategy are discussed whenever
relevant.

\subsection{The model spectrum and its parameters}

The equation for the model SSP spectrum fitted by {\sc starlight} is

\begin{equation}
\label{eq:SinglePopSpectrum}
M_\lambda = x \gamma_\lambda^{SSP}(t,Z) 
           10^{-0.4 A_V (q_\lambda - q_{\lambda_0})}
\end{equation}

\noindent where $x$ is a scaling factor, $q_\lambda \equiv A_\lambda /
A_V$ is the reddening curve, and

\begin{equation}
\label{eq:gamma}
\gamma_\lambda^{SSP}(t,Z) = 
  \frac{L_\lambda^{SSP}(t,Z)}{L_{\lambda_0}^{SSP}(t,Z)}
  \otimes G(v_\star,\sigma_\star)
\end{equation}

\noindent gives the spectrum of an SSP of age $t$ and metallicity $Z$
normalized at $\lambda_0 = 4020$ \AA, convolved with a gaussian filter
centered at centered at velocity $v_\star$ and with dispersion
$\sigma_\star$.  The $L_\lambda^{SSP}(t,Z)$ spectra are taken directly
from the evolutionary synthesis models.

To find the SSP which best matches a given observed spectrum one may
construct a base of $N_\star = N_t \times N_Z = 46 \times 6 = 276$ SSP
model spectra.  For each base component {\sc starlight} then finds the
corresponding values of $A_V$ and $x$ which best match $O_\lambda$.
For practical reasons, the actual fits were carried out using 6 single
$Z$ bases, each containing all $N_t$ available ages. In the end one
has a table of $\chi^2$, $A_V$ and $x$ for each of the $N_Z \times
N_t$ combinations of $t$ and $Z$.

The free parameters in these fits are $t$, $Z$, $A_V$ and $x$. Of
these, $t$ and $Z$ are the clearly the more fundamental ones. $A_V$
also has some astrophysical value, but the scaling factor $x$ is just
a technical parameter, with no physical relevance (see
\ref{sec:ScalingFactor}).  Note that $t$ and $Z$ can only assume the
values in the base, while $A_V$ and $x$ are continuous.

\subsubsection{Extinction}
\label{sec:Extinction}

Unlike the method of K08, which concentrates on fitting absorption
lines by modeling the continuum with a high order polynomiun, or the
``Continuum-normalized'' fits of Wolf et al.\ (2007), our spectral
fits do take the continuum shape into account, so dust effects must be
considered. Considering the difficulty in defining the continuum
around the Balmer jump and the 4000 \AA\ break, fits to the full
non-rectified spectra seem more advisable. More importantly, unless
there are calibration problems with the data or the models, it is
obviously advantageous to take the full spectral information into
account.

Extinction is dealt with assuming a simple foreground screen model
(equation \ref{eq:SinglePopSpectrum}), unrealistic for galaxies but Ok
for SCs (at least for the ages of our clusters).  SMC, LMC and
Galactic clusters were fitted with reddening curves appropriate to
each of these galaxies (Gordon et al.\ 2003; Cardelli, Clayton \&
Mathis 1989).

The V-band extinction is a free parameter in our fits, but a priori
limits can be set which may, at least in principle, help constraining
the fits.  The extinction in the LMC and SMC is low. Bica \& Alloin
(1986) give a global extinction of $E{(B-V)} \leq 0.06$ and 0.03 for
the SCs in LMC and SMC, respectively, and between 0.0 and 0.09 for the
GCs in our sample. We have also compiled LMC extinction values using
the web tool described by Zaritsky et al.\ (2004), which gives $A_V$
values along the line-of-sight to stars within a search radius of the
target coordinates. At the positions of the LMC SCs, we obtain $A_V$
ranging from 0.3 to 0.7, with a large dispersion: $\sigma(A_V) =
0.3$--0.6. These values are much larger than those given by Bica \&
Alloin (1986), but Zaritsky et al. explain that, given the highly
non-gaussian distribution of $A_V$ values, this method can only
provide a rough estimation of the extinction.  As a whole, $A_V \le 1$
is a safe upper limit for all SCs studied here.

We also impose the physical limit $A_V \ge 0$. While natural, this
choice deserves a comment. Imposing $A_V \ge 0$ prevents {\sc
starlight} from compensating for possible flux calibration problems in
the stellar libraries. Such problems are known to exist with STELIB
(Le Borgne et al.\ 2003).  Comparing stars in common between MILES and
STELIB, S\'anchez-Bl\'azquez et al.\ (2006) find the latter to be slightly
too red.  This is the reason why $\sim$ dustless galaxies are somewhat
better fitted with slightly negative values of $A_V$ (of the order of
$-0.1$ mag) when the STELIB-based Bruzual \& Charlot (2003) models are
used, while with MILES this problem goes away (Gomes 2009; Cid
Fernandes et al.\ 2009).

\subsubsection{Scaling factor}
\label{sec:ScalingFactor}

Since both model and observed spectra are normalized at $\lambda_0$,
the scaling factor $x$ in equation (\ref{eq:SinglePopSpectrum}) would
seem an unnecessary parameter. This is not strictly true.  First,
models and data are normalized in slightly different ways: While the
SSP models are normalized exactly at $\lambda_0 = 4020$ \AA, the
observed spectrum $O_\lambda$ is actually normalized by the median
flux in the 4010--4060 \AA\ window---{\sc starlight} uses this trick
to circumvent eventual problems in the $\lambda_0$ pixel. Second,
after convolution with the kinematical filter $G$, the
$\gamma_\lambda^{SSP}$ spectra are not exactly $= 1$ at $\lambda_0$
(see equation \ref{eq:gamma}). Regardless of these technicalities,
forcing a $M_{\lambda_0} = O_{\lambda_0}$ normalization would imply
treating $\lambda_0$ differently from other $\lambda$'s, ultimately
making the whole fit formally dependent on the arbitrary choice for
$\lambda_0$.  For these reasons, the scaling factor must be considered
a free parameter, even though one does not expect large departures
from $x = 1$.

\subsubsection{Kinematical parameters}
\label{KinematicalParameters}

Formally, $v_\star$ and $\sigma_\star$ could also be considered free
parameters.  However, these are {\em not} varied during the
single-population fits. Instead, they are fixed at the values
determined during the initial {\em multi-population} fit, where the
spectrum is computed from

\begin{equation}
\label{eq:MultiPopSpectrum}
M_\lambda = \sum_{j=1}^{N_\star} x_j \gamma_\lambda^{SSP}(t_j,Z_j) 
    \times 10^{-0.4 A_V (q_\lambda - q_{\lambda_0})}  
\end{equation}

\noindent instead of equation (\ref{eq:SinglePopSpectrum}). As
expected, we always obtain $v_\star$ within a few km$\,$s$^{-1}$ of
zero, since the spectra are in the rest-frame. Hence, $v_\star$ should
not be considered as a truly free parameter.
The same applies to $\sigma_\star$, which, for the best fit models, is
always close to the value corresponding to the difference between
spectral resolution of the models and the data---the spectral
resolution of the data analyzed here is $\sigma_{inst} \sim 170$
km$\,$s$^{-1}$, much larger than the actual velocity dispersion of
SCs, so that no useful kinematical information can be derived.  We
therefore do not count $\sigma_\star$ as a free parameter either. 

{\sc starlight} allows the user to fix $\sigma_\star$ at its
expected value ($\sim \sigma_{inst}$, in our case), and this would be
a valid strategy for the SC data analyzed here.  Given our goal of
mimicking inasmuch as possible the way that spectral synthesis is
carried out with galaxy data, we opted not to fix $\sigma_\star$. For
SC spectra, this can be seen as conservative choice, since
$\sigma_\star$ can be used to compensate for a metallicity mismatch
(Koleva et al.\ 2009). Increasing $\sigma_\star$ has the effect of
decreasing line depths, which for metal lines can mimic the effect of
decreasing metallicity, and vice-versa, so fitting a low $Z$ system
with a higher $Z$ model and free $\sigma_\star$ will yield a $\chi^2$
not as bad as with a fixed $\sigma_\star$. This effect is in fact
detected in our fits. We find that the returned $\sigma_\star$ values
tend to increase with $Z$. Typical values span the 150 to 200
km$\,$s$^{-1}$ range, bracketing $\sigma_{inst}$.

Overall, however, the impact of our choice to let $\sigma_\star$ free
is insignificant. Comparing SC parameters derived with fits with fixed
and free $\sigma_\star$ leads to typical differences of 0.07 or less
for $\log t$, $\log Z$ and $A_V$. We are clearly not yet in a position
to aim this level of accuracy. For instance, as shown in Paper II,
such differences are much smaller than those resulting from the use of
different evolutionary synthesis models.

\subsection{Error spectrum}
\label{sec:Errors}

The figure of merit used by {\sc starlight} fits is a standard

\begin{equation}
\label{eq:chi2Def}
\chi^2 = \sum_{\lambda} \frac{(O_\lambda - M_\lambda)^2}{\epsilon_\lambda^2}
\end{equation}

\noindent where $\epsilon_\lambda$ is the error in $O_\lambda$. As is
often the case, we do not have formal values for the errors.  The
global amplitude of $\epsilon_\lambda$ is not important for the
minimization of $\chi^2$, as it does not change the shape of the
likelihood function, and, as explained below, $\chi^2$ values are
rescaled anyway.  The shape of the error spectrum, which defines
whether certain pixels are given more weight than others, is more
relevant.

Two recipes for $\epsilon_\lambda$ were explored. In the first one we
set $\epsilon_\lambda = 0.1 \overline{O_\lambda}$, ie, a flat error
spectrum with amplitude equal to 10\% of the average flux, such that
all pixels are given the same weight. The second recipe is
$\epsilon_\lambda = 0.1 O_\lambda$, which gives a larger weight to
absorption lines. We have verified that the results obtained with
these two recipes are nearly always identical, so only results for the
second recipe are presented.  When the two recipes yield different
results the best fit model is clearly inadequate anyway. This happens,
for instance, when trying to fit very metal poor SCs with models of
higher metallicity.

\subsection{$\chi^2$ re-scaling and the effective number of degrees of freedom}
\label{sec:Ndof}

The lack of proper errors implies that our $\chi^2$ values do not have
a formal statistical meaning. While this does not affect the
identification of a best model, the absolute scale of $\chi^2$ is
relevant to assess uncertainties in the derived parameters.  In cases
like this, it is common to re-scale the errors such that the best
model has a $\chi^2$ equal to the number of degrees of freedom (e.g.,
Barth et al.\ 2002; Garcia-Rissmann et al.\ 2005). Using a subscript
``ST'' to denote the $\chi^2$ value returned by {\sc starlight} and
``min'' to indicate the best model, the corrected $\chi^2$ is then
given by

\begin{equation}
\chi^2 = N_{dof} \frac{\chi^2_{ST}}{\chi^2_{ST,min}}
\end{equation}

\noindent For convenience, we define

\begin{equation}
\label{eq:Def_delta_chi2}
\delta_{\chi^2} \equiv 
   \frac{\chi^2 - \chi^2_{min}}{\chi^2_{min}} =
   \frac{\chi^2_{ST} - \chi^2_{ST,min}}{\chi^2_{ST,min}}
\end{equation}

\noindent such that changes in $\chi^2$ scale with $\delta_{\chi^2}$:

\begin{equation}
\label{eq:DeltaChi2}
\Delta \chi^2 = \chi^2 - \chi^2_{min} = N_{dof} \delta_{\chi^2}
\end{equation}

In these equations $N_{dof} = N_\lambda - N_{par}$, i.e., the number
of data points minus the number of free parameters. $N_{par}$ is well
defined: There are 4 free parameters ($t$, $Z$, $A_V$ and $x$). But
what should be used for $N_\lambda$? In other words: {\em How many
observables are we actually fitting?}

Our fits are performed with a $\Delta\lambda = 1$ \AA\ sampling, so
there are 951 data points between 3650 and 4600 \AA.  However, the
observed spectra are highly oversampled, so the number of actually
independent data points is $\ll 951$.  Counting only points separated
by 1 FWHM$_{inst} = 5.7$ \AA\ would yield $N_\lambda = 167$, but this
is still an overestimate, given the heavy overlap in the instrumental
profiles for a spectral distance of just 1 FWHM$_{inst}$. To be safe,
only points separated by $6 \sigma_{inst} = 14.5$ \AA\ will be counted
as independent, such that the overlap in instrumental profiles occurs
beyond $\pm 3 \sigma_{inst}$, being therefore insignificant. This
somewhat subjective but clearly conservative choice yields $N_\lambda
= 65$, and hence $N_{dof} = 61$. (These values are slight smaller in a
few cases due to masked windows.)  This recipe only affects the
numerical interpretation of the statistical confidence associated to a
given value of $\Delta \chi^2$, which can be straightforwardly
recomputed for any other choice of $N_{dof}$.

In the presentation of the spectral fits two other figures of merit
are used: the rms of the $R_\lambda = O_\lambda - M_\lambda$ residual
and the mean value of $|R_\lambda| / O_\lambda$, denoted by
$\overline{\Delta}$. These numerically similar indices give an
easy-to-interpret measure of the quality of the fit.

\section{Spectral fitting: Results}
\label{sec:SpecFitResults}

The method outlined above was applied to the 27 SC spectra described
in Section \ref{sec:Data}, after basic pre-processing steps, such as
resampling to $\Delta \lambda = 1$ \AA, and masking bad pixels. This
section presents the results obtained.  

First we illustrate our methodology with fixed $Z$ fits (Section
\ref{sec:SingleZFits}) and discuss the meaning of multi-SSPs results
for SCs (Section \ref{sec:mSSP_Fits}).  Then, in Section
\ref{sec:MultiZFits}, we inspect how much the fits constrain $t$, $Z$
and $A_V$ (a task which will be readdressed more rigorously in Section
\ref{sec:Bayes}), and investigate differences in spectral fits
obtained with different $Z$'s.  Third, we map covariances
(``degeneracies'') between parameters (Section \ref{sec:chi2Maps}).
For didactic purposes, the LMC clusters NGC 2010 and NGC 2210 are taken
as examples.  Section \ref{sec:AllFitsResults} presents results for
all SCs.

\subsection{Fixed $Z$ analysis}
\label{sec:SingleZFits}

\begin{figure*}
\includegraphics[bb= 40 170 560 460,width=0.9\textwidth]{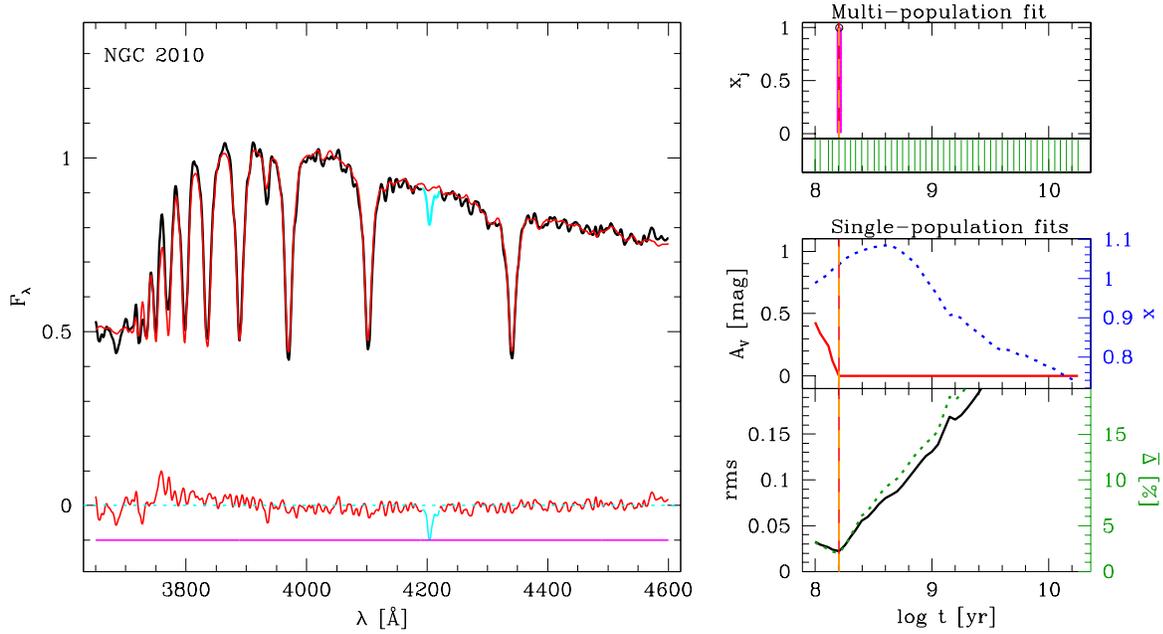}
\caption{{\em Left:} Observed (black) and best-fit (red) spectra of
NGC 2010 for $Z = Z_\odot$ V00s models. The residual spectrum is shown
at the bottom.  All spectra are in units of the flux at $\lambda_0 =
4020$ \AA. The ``absorption-line'' (marked in cyan) at
$\sim 4205$ \AA\ is actually a defect in the original spectrum,
masked from the fits.  {\em Top right:} Results of the multi-SSP
fit. The ``bar code'' shows the 46 ages in the base. {\em Middle
right:} Best fit $A_V$ and $x$ for single-SSP fits as a function of
age. Solid (red) and dashed (orange) vertical lines (which coincide in
this example) mark the best single-SSP age and mean $\log t$ in the
multi-SSP fit. {\em Bottom right:} rms and $\overline{\Delta}$ figures
of merit as a function of age. The best fit model for this $Z$ has an
age $\log t = 8.20$ and $A_V = 0.00$, producing an rms of $0.020$ and
$\overline{\Delta} = 2.0$ percent. Fits with $\log Z/Z_\odot = +0.2$
($1.5 Z_\odot$) yield a miniscule 1\% improvement in $\chi^2$, and
$\log t = 8.11$, $A_V = 0.11$.}
\label{fig:ExampleFit1} 
\end{figure*}

\begin{figure*}
\includegraphics[bb= 40 170 560 460,width=0.9\textwidth]{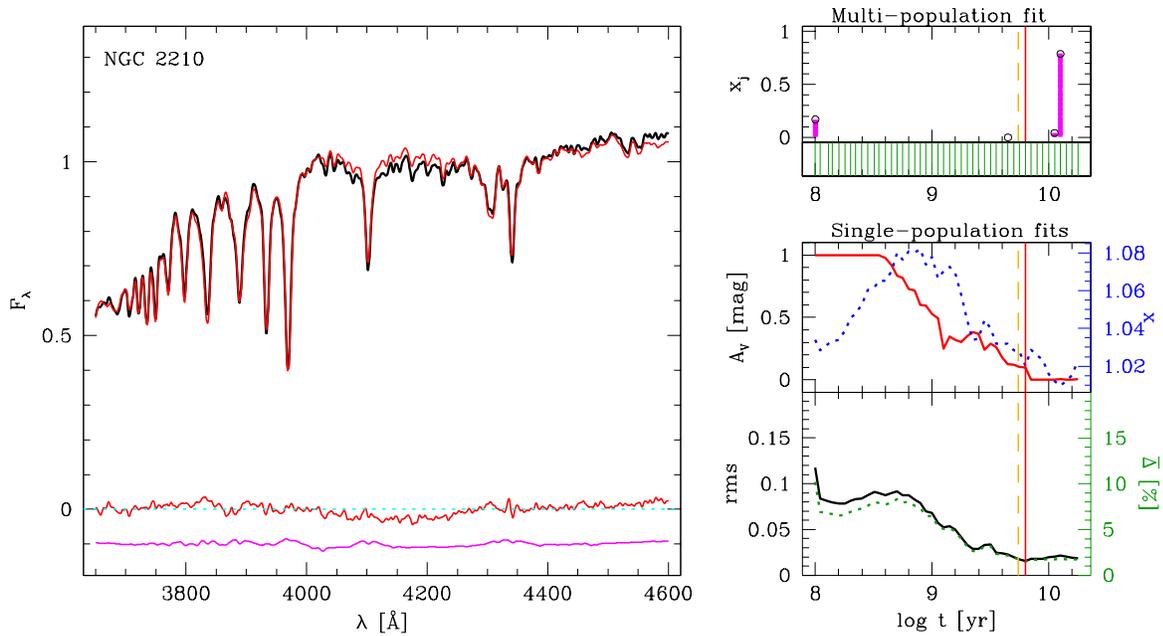}
\caption{As Fig.~\ref{fig:ExampleFit1}, but for NGC 2210 and $\log
Z/Z_\odot = -1.7$ ($0.02 Z_\odot$). Notice that, unlike for NGC 2010,
the multi-SSP fit splits into 2 widely separated ages.  The spectral
difference between the multi and single population fits  is
shown as a magenta line in the left panel, off-set by -0.1 for
clarity.}
\label{fig:ExampleFit2} 
\end{figure*}

Fig.~\ref{fig:ExampleFit1} shows the fit to NGC 2010 obtained with $Z$
fixed at $Z_\odot$.  Insignificantly better ($\delta_{\chi^2} = 0.01$)
fits are obtained with the $1.5 Z_\odot$ models. The left panel shows
the observed, model, and residual spectra. The best match is obtained
for $\log t = 8.20$, $A_V = 0.00$ and $x = 1.03$. The middle right
panel shows the runs of $A_V$ and $x$ as a function of $t$, while the
bottom one shows the two indicators of fit quality discussed at the
end of Section \ref{sec:Ndof}. The largest differences between
$O_\lambda$ and $M_\lambda$ are found in the continuum between higher
order Balmer lines and at the bottom of the CaII K line, but these are
relatively small deviations. Overall, the spectral fit is very good
both in the continuum and absorption lines.

Fig.~\ref{fig:ExampleFit2} shows the results for NGC 2210 and $\log
Z/Z_\odot = -1.7$.  This is the $Z$ which produces the smallest
$\chi^2$ among all 6 $Z$'s in the V00s models. Since $\log Z/Z_\odot =
-1.7$ is also the smallest one in the grid, it is possible that models
with even lower $Z$ would provide better fits.  In this case
differences between $O_\lambda$ and $M_\lambda$ concentrate around
4200 \AA.  The best fit is achieved for $\log t = 9.80$ and $A_V =
0.10$.  Fits for younger ages require a larger extinction to
compensate for the bluer predicted colors, but even then the quality
of the fit deteriorates quickly as one moves away from the best
$t$. Notice also that $A_V$ saturates at the imposed limits as one
moves away from the best model.

\subsection{The meaning of multi component fits}
\label{sec:mSSP_Fits}

Accepting that SCs are single population systems, multi-SSP fits are
in principle not of direct interest. Still, even if only out of
curiosity, lets have a look at them.

The top right panel in Fig.\ \ref{fig:ExampleFit1} shows how the light
at $\lambda_0$ is spread among SSPs of different ages in the multi-SSP
{\sc starlight} fit of NGC 2010. In this particular example, the
multi-SSP fit is {\em identical} to the single-SSP one. In other
words, given total freedom to mix 46 different base elements, {\sc
starlight} preferred to use only one. In the $Z = 1.5 Z_\odot$ fits
(not shown), the multi-SSP fit splits into two similar parts,
corresponding to populations within 0.1 dex of the best-fit single SSP
age. The fit is thus well focused, such that multi and single SSP fits
are almost identical.

The situation is very different in NGC 2210 (Fig.\
\ref{fig:ExampleFit2}).  The multi-SSP fit in this case shows two
dominant components with ages over 2 dex apart: One with $\log t =
10.10$, responsible for about 80\% of the light, and other at the
youngest age $\log t = 8.00$, accounting for the rest.  Unlike in
Fig.~\ref{fig:ExampleFit1}, the difference between the multi and
single-SSP spectra (shown in the bottom of the left panel) is now
noticeable. Accordingly, the multi-SSP fit yields a $\chi^2$ which is
27\% better than the single-SSP one (a difference which is of only 5\%
for NGC 2010).

A plausible interpretation of this ``old plus a bit of young''
population mixture, which happens in at least two other clusters in
our sample, is that the SSP models lack old blue stars. This same
problem was identified by K08 in a different sample. They found that
clusters with a blue horizontal branch (HB) are better fitted adding a
set of stars in the $T_{\rm eff} = 6000$ to 20000 K range to a pure
SSP. They also find that this {\em ad hoc}, but physically reasonable,
recipe produces ages in better agreement with those derived from CMD
studies.  We will return to this issue in Section
\ref{sec:Summary}. For the moment, we note that the example of NGC
2210 fully corroborates this interpretation: The fiducial age adopted
by LR03 is $\log t = 10.1$, identical to that of the oldest population
in our multi-SSP fits, but 0.3 dex older than the one obtained with a
single-SSP fit ($\log t = 9.80$).

These results show that multi-SSP fits provide an useful empirical
measure of the adequacy of single-SSP fits.  Six quantities are used
to report results of the multi-SSP fits: The mean ($\overline{\log
t}_m$) and standard deviation ($\sigma_m$) of $\log t$ (computed
directly from the population vector $\vec{x}$ using equations 2 and 4
of Cid Fernandes et al.\ 2005), the age of the dominant population
($\log T_m$) and its corresponding percentage light fraction ($f_m$),
plus the mean relative residual ($\overline{\Delta}_m$), and the
difference in $\chi^2$ between the multi and single SSP fits
($\delta_m$), defined as in equation (\ref{eq:Def_delta_chi2}). These
quantities are tabulated in Table \ref{tab:Results_BVM00s}, along with
other results.  For NGC 2010 (Fig.\ \ref{fig:ExampleFit1}), for
instance, we find $\overline{\log t}_{m} = 8.12$, $\sigma_{m} = 0.13$,
and $\delta_m = 5\%$, confirming a focused single SSP solution. On the
other hand, for NGC 2210 (Fig.\ \ref{fig:ExampleFit2}),
$\overline{\log t}_{m} = 9.74$, $\log T_{m} = 10.10$, $\sigma_{m} =
0.79$, and $\delta_m = 27\%$, signalling the spread in ages discussed
above. Section \ref{sec:Summary} elaborates on how these numbers can
be used to identify suspicious fits.


\subsection{Fits with different metallicities}
\label{sec:MultiZFits}

\begin{figure*}
\includegraphics[bb= 40 170 580 530,width=0.85\textwidth]{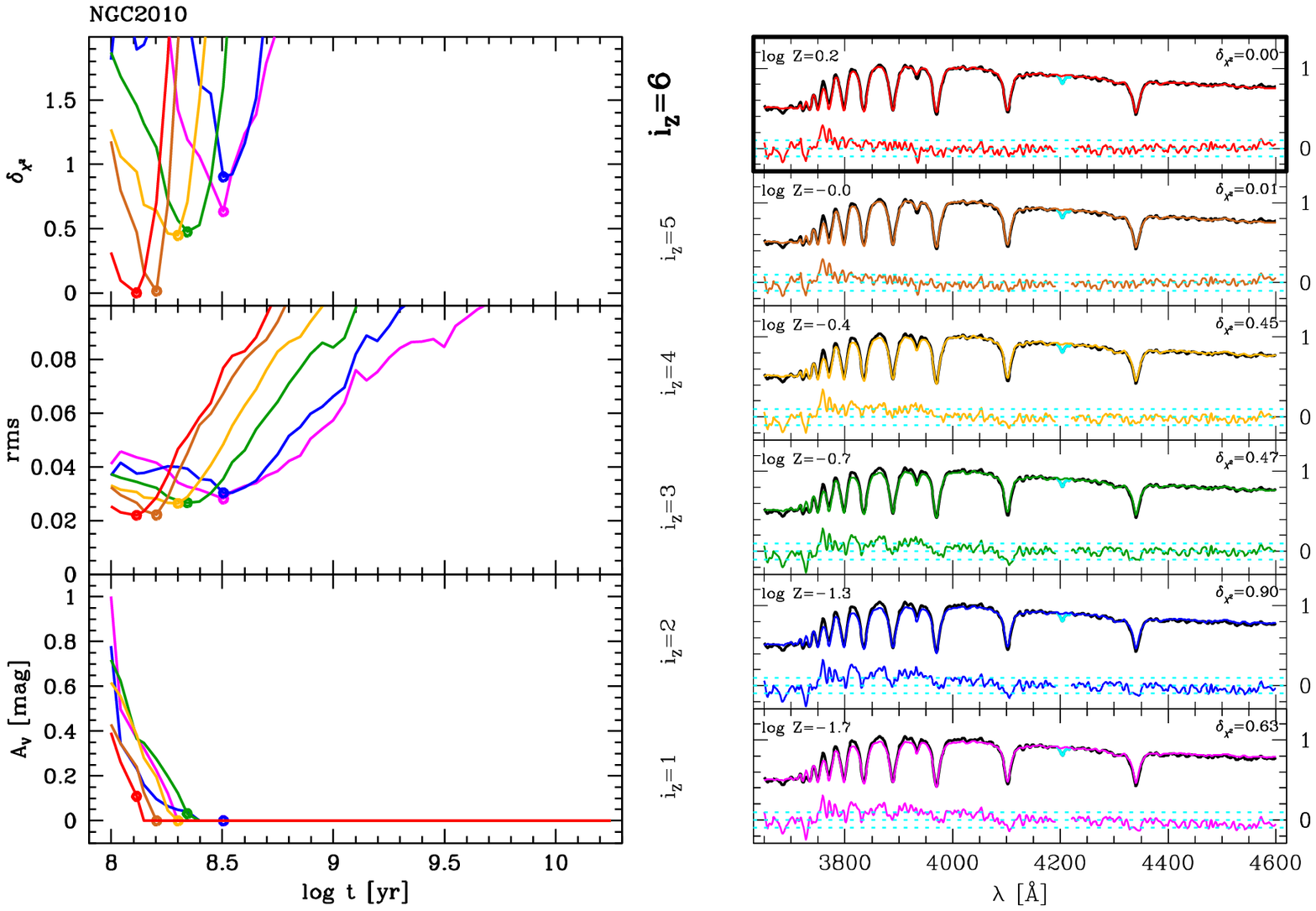}
\caption{Results of spectral fits of NGC 2010 with 6 different
metallicities. {\em Left:} Run of best-fit $A_V$ (bottom), rms
(middle) and $\delta_{\chi^2}$ (top) with age. Different lines
correspond to different $Z$'s: magenta, blue, green, orange, brown and
red correspond to $\log Z/Z_\odot = -1.7$, -1.3, -0.7, -0.4, 0, and
+0.2, respectively. Circles mark the best-fit model for each $Z$.
{\em Right:} Best fit SSP spectra for all 6 different $Z$'s. The
observed model is in black and the model and residual spectra are
plotted according to the color-scheme above. Note: The $O_\lambda -
M_\lambda$ residual spectra are multiplied by 3 for clarity. Dashed
horizontal lines at $y = -0.1$, 0 and $+0.1$ are drawn to facilitate
comparisons. The best fit model ($\log Z/Z_\odot = +0.2$ in this
case) is marked by a thicker window frame.
}
\label{fig:AVrmsAndFits_n2010}
\end{figure*}

\begin{figure*}
\includegraphics[bb= 40 170 580 530,width=0.85\textwidth]{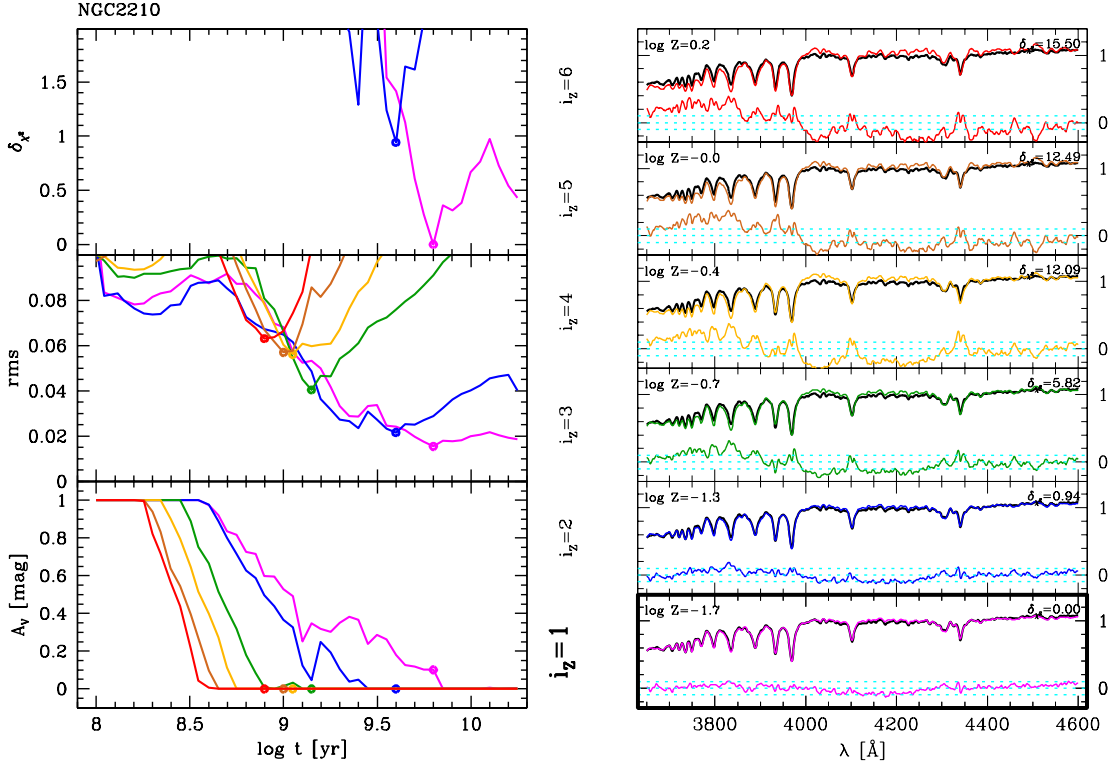}
\caption{As Fig.\ \ref{fig:AVrmsAndFits_n2010} but for NGC 2210.}
\label{fig:AVrmsAndFits_n2210}
\end{figure*}

The example fits in Figs.~\ref{fig:ExampleFit1} and
\ref{fig:ExampleFit2} have fixed $Z$. In
Figs.~\ref{fig:AVrmsAndFits_n2010} and \ref{fig:AVrmsAndFits_n2210} we
show how things change for different $Z$'s.  The bottom-left panels
show the best-fit $A_V$ as a function of age for all 6
metallicities. The middle panel show the rms of the single-SSP
spectral fits, while the top panel shows $\delta_{\chi^2}$. Curves are
color-coded according to $Z$, and a circle marks the location of the
best model for the corresponding $Z$.  Panels in the right show the
best spectral fit achieved for each of the 6 metallicities. Residual
spectra shown in these panels were {\em multiplied by 3} for clarity.

Fig.~\ref{fig:AVrmsAndFits_n2010} shows results for NGC 2010.  Among
all $6 \times 46 = 276$ SSPs, the one with $\log Z/Z_\odot = +0.2$ and
$\log t = 8.11$ is the one which produces the best fit. Accordingly,
this model has $\delta_{\chi^2} = 0$. As already noted, the next best
$Z$ is solar, for which $\delta_{\chi^2} = 0.01$, i.e., its $\chi^2$
is just 1\% worse than the best one! Visual inspection of the
corresponding spectra confirms that these two solutions are indeed
indistinguishable. Other metallicities produce visibly worse fits,
suggesting that $Z$ is relatively well constrained to $\ge Z_\odot$
for this SC.

We note in passing that neither the rms (middle left panel) nor
$\overline{\Delta}$ (not plotted) provide a strong numerical
discrimination of different models. As can be seen in
Fig.~\ref{fig:AVrmsAndFits_n2010}, the best models for different $Z$'s
all yield numerically similar values for these figures of merit, even
when the fits are visibly worse. Even if formally equivalent, the
$\delta_{\chi^2}$ index is much more useful in this sense.

Results for NGC 2210 are shown in Fig.\
\ref{fig:AVrmsAndFits_n2010}. In this case, there is a well defined
metallicity: $\log Z/Z_\odot = -1.7$. The next best model ($\log
Z/Z_\odot = -1.3$) has a nearly twice as bad $\chi^2$
($\delta_{\chi^2} = 0.94$). Fits for higher $Z$'s are so much worse
that they fall off the top right panel.

\subsection{Age-metallicity-extinction degeneracies}
\label{sec:chi2Maps}

\begin{figure}
\includegraphics[bb= 40 220 210 690,width=0.45\textwidth]{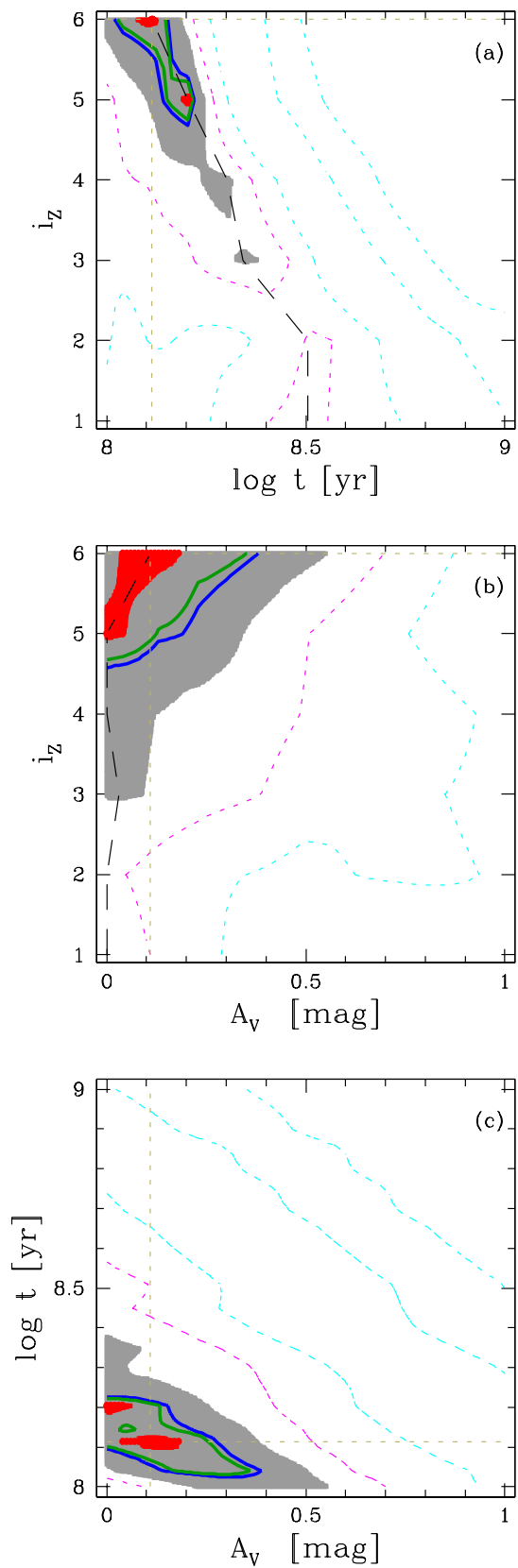}
\caption{$\delta_{\chi^2}$ contours in the age-metallicity-extinction
space for NGC 2010. The magenta line correspond to $\delta_{\chi^2} =
1$ and the gray shaded area marks the $\delta_{\chi^2} \le 0.5$
region. Blue, green and red correspond to 3, 2 and 1 sigma confidence
regions, respectively. Best-fit values are marked with dotted
lines. In the top panels a dashed line connects the best models for
different $Z$'s.}
\label{fig:All3Contours_NGC2010}
\end{figure}

\begin{figure}
\includegraphics[bb= 40 220 210 690,width=0.45\textwidth]{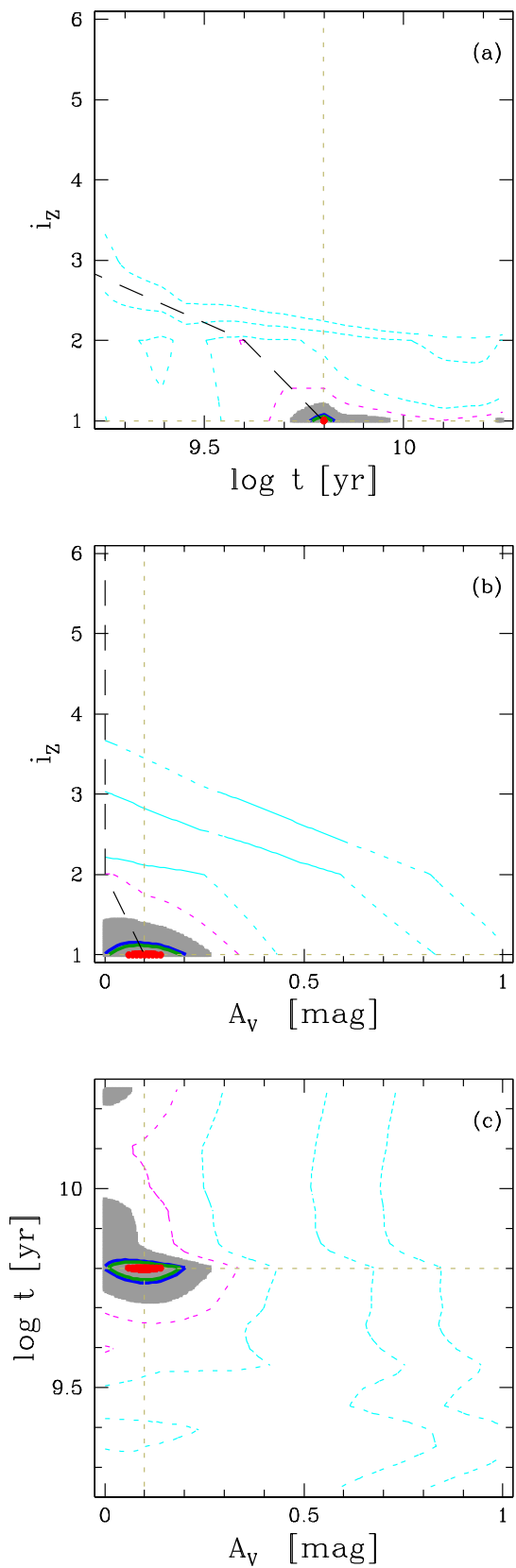}
\caption{As Fig.\ \ref{fig:All3Contours_NGC2010} but for NGC 2210.}
\label{fig:All3Contours_NGC2210}
\end{figure}

The plots above already show known couplings between parameters. The
best way to visualize such covariances is through bidimensional
$\Delta\chi^2$ maps.

Fig.~\ref{fig:All3Contours_NGC2010} shows maps of $\delta_{\chi^2}$
in the three possible projections of the $t$-$Z$-$A_V$ space for NGC
2010. Notice that the $Z$ scale is not in physical units. Instead, we
prefer to represent each of the 6 $Z$'s by a $i_Z = 1$ to 6
metallicity index, a cosmetic choice which serves to emphasize the
coarseness of the $Z$-grid.  Dotted contours are for $\delta_{\chi^2}
\ge 1$, the inner one ($\delta_{\chi^2} = 1$) being drawn in magenta.
The gray-shaded area maps the $\delta_{\chi^2} \le 0.5$ zone, while
the inner 3 contours mark the formal 3, 2 and 1 sigma ranges for our
choice of $N_{dof}$ (see equation \ref{eq:DeltaChi2}).

The well known trade-off between age and metallicity is evident in
Fig.~\ref{fig:All3Contours_NGC2010}a. Notice also that solar and
over-solar models are equally good, as noted before in Fig.\
\ref{fig:AVrmsAndFits_n2010}, where best fit spectra for each $Z$ were
shown. These two $Z$'s fall within the 1 sigma confidence region. In
the case of NGC 2210, the ``age-metallicity degeneracy'' is not so
evident because the best-fit metallicity is at the border of the
$Z$-grid (Fig.~\ref{fig:All3Contours_NGC2210}a).

As $t$ and $Z$ assume only the discrete values dictated by the set of
models used, such $\Delta \chi^2(t,Z)$ maps can be constructed
directly from the output of the {\sc starlight} fits. However, {\sc
starlight} only computes the best $A_V$ (and $x$) for each $(t,Z)$
pair, so some extra coding is needed to produce maps involving
$A_V$. We have thus written a complementary code which, for each $t$
and $Z$, forces fits using $A_V$ values in a fine grid from 0 to
1. These fits have a single free parameter, $x$, whose optimal value
is computed analytically from $\partial \chi^2 / \partial x = 0$. The
resulting tables of $\chi^2$ as a function of $t$, $Z$ and $A_V$ can
then be projected to $\Delta\chi^2(t,A_V)$ and $\Delta\chi^2(Z,A_V)$
maps.

As colors redden with increasing age and metallicity, one expects
negative covariances between $A_V$ and both $t$ and $Z$.  Such an
``age-extinction degeneracy'' is indeed observed in
Fig.~\ref{fig:All3Contours_NGC2010}c (NGC 2010). The effect is also
present, but much weaker, in NGC 2210 (Fig.\
\ref{fig:All3Contours_NGC2210}c) where the trend is only noticeable in
the $\delta_{\chi^2} \ge 0.5$ contours.

For many clusters, contours in the $A_V$-$Z$ plane show the expected
``metallicity-extinction degeneracy'', but in these two examples it is
not so clear. A hint of the expected effect is present in NGC 2210
(Fig.\ \ref{fig:All3Contours_NGC2210}b), visible mainly from the outer
contours, since the best fit values of $Z$ and $A_V$ are at a corner
of the grid.  In NGC 2010, however, the trend is opposite to the
expected one, with contours showing positive covariance (Fig.\
\ref{fig:All3Contours_NGC2010}b). In this case, the blueing of the
continuum due to the decrease in $t$ as $Z$ increases (Fig.\
\ref{fig:All3Contours_NGC2010}a) is larger than the reddening caused
by the change in $Z$, explaining why $A_V$ increases with $Z$.  This
example serves as a reminder that the quantitative effects of the
combination of $t$-$Z$-$A_V$ ``degeneracies'' can be more subtle than
those expected on simple qualitative grounds.

These maps show that the $A_V \le 1$ prior has very little effect.
This happens because the extinction in our SCs is genuinely small.
For the same reason, the physical $A_V \ge 0$ limit has more effect in
constraining $A_V$, as seen, for instance, in the contours in Fig.\
\ref{fig:All3Contours_NGC2010}b.

Before closing, we note that, as said in Section \ref{sec:Ndof}, the
statistical interpretation of these surfaces can be adapted to other
choices for $N_{dof}$. For instance, the $\delta_{\chi^2} = 0.5$
shaded areas in Figs.\ \ref{fig:All3Contours_NGC2010} and
\ref{fig:All3Contours_NGC2210} would correspond to 3 sigma contours
($\Delta \chi^2 = 11.8$ for a 2D projection) for $N_{dof} = 24$,
equivalent to considering every 34 \AA\ of a spectrum as an
independent datum.

\subsection{Results for all clusters}
\label{sec:AllFitsResults}

\begin{figure*}
\includegraphics[bb= 40 170 580 710,width=0.95\textwidth]{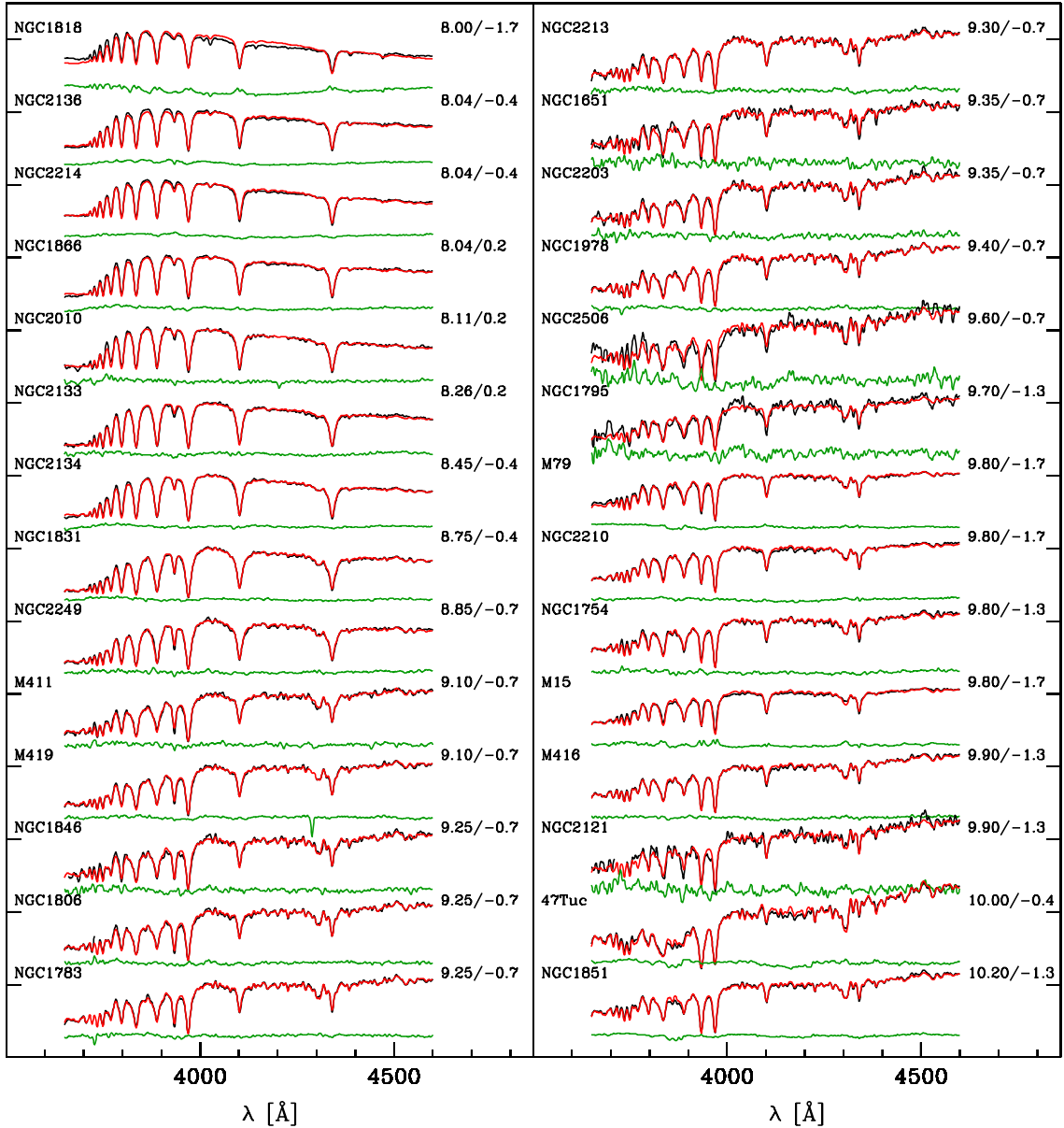}
\caption{All spectral fits, sorted by best-fit age. Observed and best
fit single SSP spectra are shown in black and red, respectively, with
the corresponding residual shown in green. Labels to the right of each
spectrum list the best values of $\log t$ and $\log Z/Z_\odot$. All
spectra are normalized at $\lambda_0 = 4020$ \AA, and shifted
vertically by 1 ($=$ one tick mark) from one another.}
\label{fig:All_BVM00s_BestFits}
\end{figure*}

\begin{table*}
\centering
\begin{tabular}{@{}lrrrrrrrrrrrr@{}} \hline
        &
\multicolumn{4}{c}{Single population fits} &
\multicolumn{5}{c}{Multi population fits} &
\multicolumn{3}{c}{Bayesian estimates} \\ \hline
Cluster   &
$\log t$  &
$\log Z/Z_\odot$  &
$A_V$     &
$\overline{\Delta}$ &
$\delta_{m}$   & 
$\overline{\log t}_{m}$ &
$\sigma_{m}$  & 
$\overline{\Delta}_{m}$ &
$\log T_{m}$ (\%) &
$\log t$ &
$\log Z/Z_\odot$  & 
$A_V$     \\  
(1)  &
(2)  &
(3)  &
(4)  &
(5)  &
(6)  &
(7)  &
(8)  &
(9)  &
(10) &
(11) &
(12) &
(13) \\ \hline
%
M   0411 &  9.10 & -0.68 & 0.48 &  2.3 &  0.05 &  9.10 &  0.35 &  2.2 &  8.95  ($ 43$)  &  9.13 $\pm$  0.10 & -0.68 $\pm$  0.20 & 0.41 $\pm$ 0.11 \\  
M   0416 &  9.90 & -1.28 & 0.22 &  1.4 &  0.00 &  9.89 &  0.06 &  1.4 &  9.90  ($ 98$)  &  9.90 $\pm$  0.01 & -1.28 $\pm$  0.00 & 0.22 $\pm$ 0.02 \\  
M   0419 &  9.10 & -0.68 & 0.46 &  1.6 &  0.10 &  9.12 &  0.11 &  1.5 &  9.20  ($ 41$)  &  9.10 $\pm$  0.00 & -0.67 $\pm$  0.06 & 0.46 $\pm$ 0.05 \\  
NGC 1651 &  9.35 & -0.68 & 0.00 &  4.0 &  0.06 &  9.37 &  0.29 &  3.9 &  9.05  ($ 44$)  &  9.58 $\pm$  0.19 & -1.10 $\pm$  0.34 & 0.36 $\pm$ 0.19 \\  
NGC 1754 &  9.80 & -1.28 & 0.00 &  1.9 &  0.13 &  9.74 &  0.28 &  1.8 &  9.90  ($ 76$)  &  9.81 $\pm$  0.03 & -1.28 $\pm$  0.04 & 0.02 $\pm$ 0.02 \\  
NGC 1783 &  9.25 & -0.68 & 0.24 &  1.6 &  0.03 &  9.22 &  0.27 &  1.6 &  9.05  ($ 53$)  &  9.25 $\pm$  0.06 & -0.69 $\pm$  0.13 & 0.26 $\pm$ 0.10 \\  
NGC 1795 &  9.70 & -1.28 & 0.39 &  5.2 &  0.02 &  9.66 &  0.35 &  5.1 &  9.90  ($ 67$)  &  9.66 $\pm$  0.16 & -1.21 $\pm$  0.25 & 0.38 $\pm$ 0.17 \\  
NGC 1806 &  9.25 & -0.68 & 0.37 &  1.7 &  0.12 &  9.25 &  0.19 &  1.6 &  9.15  ($ 46$)  &  9.26 $\pm$  0.02 & -0.67 $\pm$  0.03 & 0.34 $\pm$ 0.08 \\  
NGC 1818 &  8.00 & -1.68 & 0.00 &  3.9 &  0.00 &  8.02 &  0.20 &  3.9 &  8.00  ($ 99$)  &  8.00 $\pm$  0.00 & -1.68 $\pm$  0.00 & 0.02 $\pm$ 0.02 \\  
NGC 1831 &  8.75 & -0.38 & 0.00 &  1.7 &  0.16 &  8.72 &  0.16 &  1.6 &  8.75  ($ 94$)  &  8.73 $\pm$  0.03 & -0.40 $\pm$  0.08 & 0.03 $\pm$ 0.04 \\  
NGC 1846 &  9.25 & -0.68 & 0.21 &  2.8 &  0.00 &  9.24 &  0.15 &  2.8 &  9.25  ($ 94$)  &  9.23 $\pm$  0.07 & -0.70 $\pm$  0.13 & 0.27 $\pm$ 0.12 \\  
NGC 1866 &  8.04 &  0.20 & 0.55 &  1.9 &  0.04 &  8.08 &  0.12 &  1.9 &  8.00  ($ 70$)  &  8.05 $\pm$  0.03 &  0.19 $\pm$  0.07 & 0.53 $\pm$ 0.07 \\  
NGC 1978 &  9.40 & -0.68 & 0.09 &  1.7 &  0.12 &  9.42 &  0.41 &  1.6 &  9.60  ($ 82$)  &  9.42 $\pm$  0.11 & -0.69 $\pm$  0.10 & 0.11 $\pm$ 0.09 \\  
NGC 2010 &  8.11 &  0.20 & 0.11 &  2.0 &  0.05 &  8.12 &  0.13 &  2.0 &  8.00  ($ 54$)  &  8.14 $\pm$  0.04 &  0.14 $\pm$  0.09 & 0.09 $\pm$ 0.07 \\  
NGC 2121 &  9.90 & -1.28 & 0.71 &  4.5 &  0.00 & 10.05 &  0.17 &  4.5 &  9.90  ($ 56$)  &  9.99 $\pm$  0.19 & -1.23 $\pm$  0.17 & 0.55 $\pm$ 0.19 \\  
NGC 2133 &  8.26 &  0.20 & 0.31 &  2.1 &  0.09 &  8.32 &  0.15 &  2.0 &  8.26  ($ 77$)  &  8.27 $\pm$  0.03 &  0.11 $\pm$  0.10 & 0.33 $\pm$ 0.09 \\  
NGC 2134 &  8.45 & -0.38 & 0.03 &  1.6 &  0.00 &  8.45 &  0.00 &  1.6 &  8.45  ($100$)  &  8.41 $\pm$  0.08 & -0.29 $\pm$  0.19 & 0.09 $\pm$ 0.10 \\  
NGC 2136 &  8.04 & -0.38 & 0.54 &  2.1 &  0.05 &  8.07 &  0.16 &  2.0 &  8.00  ($ 84$)  &  8.12 $\pm$  0.11 & -0.47 $\pm$  0.19 & 0.40 $\pm$ 0.19 \\  
NGC 2203 &  9.35 & -0.68 & 0.24 &  2.8 &  0.08 &  9.42 &  0.63 &  2.6 &  8.95  ($ 27$)  &  9.35 $\pm$  0.05 & -0.69 $\pm$  0.09 & 0.25 $\pm$ 0.10 \\  
NGC 2210 &  9.80 & -1.68 & 0.10 &  1.5 &  0.27 &  9.74 &  0.79 &  1.3 & 10.10  ($ 79$)  &  9.80 $\pm$  0.00 & -1.68 $\pm$  0.00 & 0.10 $\pm$ 0.03 \\  
NGC 2213 &  9.30 & -0.68 & 0.13 &  2.1 &  0.02 &  9.32 &  0.32 &  2.1 &  9.60  ($ 37$)  &  9.30 $\pm$  0.00 & -0.68 $\pm$  0.01 & 0.13 $\pm$ 0.05 \\  
NGC 2214 &  8.04 & -0.38 & 0.00 &  1.7 &  0.00 &  8.04 &  0.01 &  1.7 &  8.04  ($ 97$)  &  8.04 $\pm$  0.01 & -0.38 $\pm$  0.01 & 0.02 $\pm$ 0.02 \\  
NGC 2249 &  8.85 & -0.68 & 0.12 &  1.8 &  0.05 &  8.82 &  0.15 &  1.7 &  8.85  ($ 72$)  &  8.84 $\pm$  0.02 & -0.67 $\pm$  0.04 & 0.15 $\pm$ 0.06 \\  
47   Tuc & 10.00 & -0.38 & 0.00 &  2.8 &  0.01 &  9.98 &  0.06 &  2.7 & 10.00  ($ 89$)  &  9.98 $\pm$  0.05 & -0.37 $\pm$  0.04 & 0.02 $\pm$ 0.02 \\  
M   0015 &  9.80 & -1.68 & 0.00 &  1.9 &  0.44 &  9.64 &  0.87 &  1.6 & 10.10  ($ 78$)  &  9.80 $\pm$  0.00 & -1.68 $\pm$  0.00 & 0.02 $\pm$ 0.02 \\  
M   0079 &  9.80 & -1.68 & 0.00 &  1.8 &  0.97 &  9.91 &  0.61 &  1.3 & 10.10  ($ 91$)  &  9.84 $\pm$  0.11 & -1.68 $\pm$  0.00 & 0.01 $\pm$ 0.01 \\  
NGC 1851 & 10.20 & -1.28 & 0.12 &  1.7 &  0.02 & 10.14 &  0.12 &  1.7 & 10.20  ($ 79$)  & 10.17 $\pm$  0.04 & -1.28 $\pm$  0.00 & 0.11 $\pm$ 0.04 \\  
\hline
\end{tabular}
\caption{Results for all clusters. Columns 2--5 correspond to the best
single SSP fits; columns 6--10 list results of the multi-SSP fits
(with $Z$ fixed and equal to that given in column 3); columns 11-13
list the Bayesian estimates of $t$, $Z$ and $A_V$.}
\label{tab:Results_BVM00s}
\end{table*}

Columns 2--5 of Table \ref{tab:Results_BVM00s} lists the best $t$, $Z$
and $A_V$ obtained for all 27 SCs in our sample.  The corresponding
value of $\overline{\Delta}$ is listed in column 6.
Fig.~\ref{fig:All_BVM00s_BestFits} shows the corresponding spectral
fits, sorted by the best-fit age.  The first cluster shown, NGC 1818,
is actually younger than 100 Myr, the smallest age available in the
V00s models, and the fact that the best-fit is found for the lowest
available $Z$ is an artificial consequence of this limitation.  Our
results for this SC should thus be ignored.  The next three SCs (NGC
2136, NGC 2214, and NGC 1866) are also close to this limit, and
results for them should be regarded with care, even though the fits
are visibly better than in the case of NGC 1818.

\section{Bayesian estimates of star cluster properties}
\label{sec:Bayes}

The fits and $\Delta \chi^2$ maps presented above give an idea of the
kind of precision in SC parameters achievable with spectral fits.  So
far, however, our assessment of the uncertainties was mostly
qualitative. This section presents a simple Bayesian formalism adapted
to the problem of SC parameter estimation.  Revised estimates of $t$,
$Z$ and $A_V$ are presented, along with their respective
uncertainties.

\subsection{Formalism}
\label{sec:BayesFormalism}

For a data set $D$ and a model with parameters $\vec{p}$, the
posterior probability of $\vec{p}$ is given by Bayes theorem:
$P(\vec{p}|D) = P(\vec{p}) P(D|\vec{p}) / P(D)$. For a $P(\vec{p}) =$
constant prior and gaussian errors, $P(\vec{p}|D) \propto
\exp[-\frac{1}{2} \chi^2(\vec{p})]$, from which estimates of any of
the individual parameters in $\vec{p}$ can be obtained through
marginalization.

Lets momentarily denote $t \equiv \log t$, $Z \equiv \log Z/Z_\odot $,
and $A \equiv A_V$. Hence, $\vec{p}$ corresponds to $(t,Z,A,x)$, while
the data $D$ correspond to an observed spectrum ($O_\lambda$) and its
errors ($\epsilon_\lambda$). The posterior probability distribution
function (PDF) of, say, the (log) age is then given by

\begin{equation}
P(t|D) = \int \int \int P(t,Z,A,x|D) \, dZ \, dA \, dx
\end{equation}

\noindent and similarly for $Z$, $A$, and $x$. 

Computing such multi-dimensional integrals is usually a cumbersome
task, but with 4 dimensions the problem is still manageable and a
simple discretization approach suffices.  To evaluate the PDF of the
parameters, one thus needs grids in $t$, $Z$, $A$ and $x$. Grids in
$t$ and $Z$ are naturally available from the SSP models, while grids in
$A$ were already computed to produce the $\Delta\chi^2(t,A_V)$ and
$\Delta\chi^2(Z,A_V)$ maps shown above. The missing $x$-grid is the
easiest to compute, as the effects upon $\chi^2$ of a change in $x$
with respect to the optimal value can be computed analytically. In this
computation we restrict $x$ to the 0.8--1.2 range, a very generous
prior, which, however, was not formally imposed in the
single-population {\sc starlight} fits presented in Section
\ref{sec:SpecFitResults}.

The $\Delta Z_j$ steps were computed as $[Z_{j+1} - Z_{j-1}]/2$
for intermediate $Z_j$'s, while the edge bins ($j = 1$ and $N_Z$) were
taken to be symmetric.  The same scheme was followed for the other
parameters, for which the sampling is much finer.

\subsection{Results}
\label{sec:BayesResults}

\begin{figure}
\includegraphics[bb= 40 170 300 710,width=0.45\textwidth]{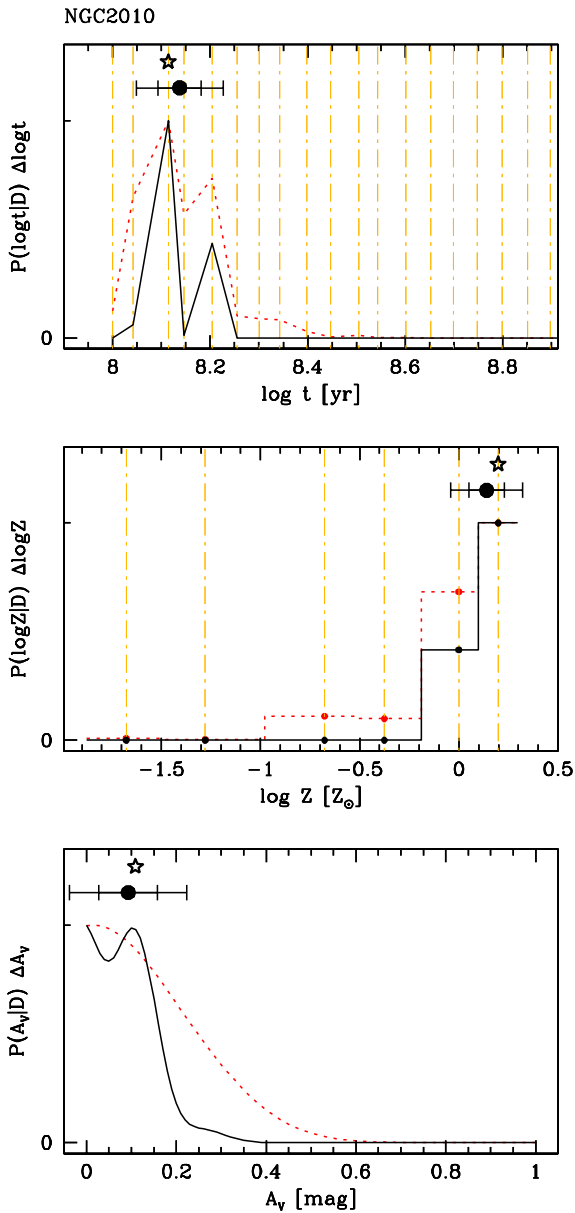}
\caption{Probability distribution functions for $t$, $Z$ and $A_V$ for
NGC 2010. Solid lines correspond to PDFs with our fiducial choice for
$N_{dof}$ (i.e., 61), whereas dotted (red) lines are for a 5 times
smaller $N_{dof}$. A star marks the best fit value. One and two sigma
error bars are also shown, centered on the mean values. In the top and
middle panels, vertical lines mark ages and metallicities available in
the set of evolutionary synthesis models used.}
\label{fig:PDFs_n2010}
\end{figure}

Results for NGC 2010 are shown in Fig.\ \ref{fig:PDFs_n2010}. In all
panels, the solid curve corresponds to PDFs computed with our fiducial
choice for $N_{dof}$ (see Section \ref{sec:Ndof}). The best fit value
is marked by a star. The average value is plotted as a circle, and
$\pm 1$ and $\pm 2$ sigma error bars (all computed directly from the
PDF) are also drawn. The correspondance with the contour plots in
Fig.~\ref{fig:All3Contours_NGC2010} is evident. The two acceptable
metallicities translate into the two peaks in $P(t)$. Two peaks are
also seen in the PDF of $A_V$, but blended.  Multi-modal PDFs and
contour plots like these occur frequently, and are basically a
by-product of the coarseness of the $Z$-grid (see Appendix
\ref{app:Appendix_ZIwork} for results obtained with $Z$-interpolated
grids).  There are, however, exceptions. NGC 2210 is an example whose
PDFs (Fig.\ \ref{fig:ZI_PDFs}) are unimodal and centered on the
best fit values, as can be guessed from
Fig.~\ref{fig:All3Contours_NGC2210}.

Just for illustrative purposes we also draw the PDF which would result
from adopting a 5 times smaller $N_{dof}$.  As expected, the PDFs
become broader, but at least in this example, this extreme choice,
which tantamounts to equalling a full spectral fit to fitting just
$N_\lambda = 16$ observables, the SC parameters are still well
constrained.

The standard way of summarizing PDFs is to compute their average and
standard deviation, as already done in Fig.\
\ref{fig:PDFs_n2010}. Results for all clusters are listed in the last
3 columns of Table \ref{tab:Results_BVM00s}. These are our official
estimates for the SC properties and their uncertainties.  Note,
however, that the table contains several entries for which
$\sigma(\log Z)$ is very small or even zero! This is obviously wrong
physically and statistically, but is a direct mathematical consequence
of the coarseness of the $Z$-grid (see Appendix
\ref{app:Appendix_ZIwork}). In such cases, NGC 2210 being an example,
the likelihood is so concentrated around one $Z$ that neighboring grid
points contain essentially no ``probability mass''. When this happens,
it is prudent to adopt something like half the $\log Z$-grid
separation as a measure of $\sigma(\log Z)$. (The mean half grid
separation is 0.2 dex in metallicity and 0.025 dex in age.)  Also, as
already noted, results for clusters too close to the age and
metallicity limits of the models should be read with caution.

Typical (sample mean) formal statistical uncertainties in $\log t$,
$\log Z/Z_\odot$ and $A_V$ are 0.06 dex, 0.09 dex and 0.08 mag. No
trend of uncertainty with mean SC properties was found.  Notice
that the typical uncertainty in $\log t$ is nearly identical to the
0.05 dex age spacing of the V00s models, so, besides a finer $Z$-grid,
a finer sampling in age would also be desirable.

Simulations which test the ability of the spectral synthesis
methodology used in this paper to recover SC parameters under
different levels of noise are described in Appendix
\ref{app:Appendix_SNwork}. The overall conclusion of these simulations
is that the method is robust. For $S/N > 30$ (as most of the data used
in this paper) one expects bayesian estimates within $< 0.1$ dex in
age, $< 0.2$ dex in metallicity and $< 0.1$ dex in $A_V$ of the
correct values.

\section{Summary and Discussion}
\label{sec:Summary}

We have presented a rigorous but straightforward methodology to fit SC
spectra with predictions from modern, publicly available high
resolution evolutionary synthesis models. The method is mostly based
on the also publicly available (but never before used for SCs)
spectral synthesis code {\sc starlight}, from which one can derive
best $t$ and $Z$ estimates, as well as likelihood maps in the $(t,Z)$
plane. Objective recipes to deal with the absence of errors and account
for spectral resolution in the definition of the effective number of
degrees of freedom were presented. A simple bayesian approach was
followed to compute formal probability distribution functions for the
parameters.  The method was applied to 27 SCs from the work of
Leonardi \& Rose (2003), modelled with publicly available SSP models
of Vazdekis et al.\ (2009) based on the MILES library, ``Padova 2000''
tracks and a Salpeter IMF.

The main goal of this paper was to setup and illustrate this
methodology, paving the way for the comparative study presented in
Paper II, where results obtained with different models are presented
and compared to independent work. This comparison is essential to map
systematic uncertainties, whereas in this paper only statistical
uncertainties were addressed.

A first general result is that the spectral fits are very good,
yielding residuals of the order of $\overline{\Delta} = 2\%$,
equivalent to a signal-to-noise ratio of 50. That, of course, does not
guarantee that the inferred parameters are correct. Insofar as $A_V$ is
concerned, our results are consistent with previous estimates, which
indicate very low extinction for SCs in this sample (e.g., Bica \&
Alloin 1986). In Paper II we show that, taking CMD determinations of
$t$ and $Z$ as a reference, spectral fitting does provide reliable
results.

Regarding the estimation of SC properties, the coarseness of the
available models in $Z$ introduces
some undesirable complications, like probability islands in the
parameter space, reflected as multimodal PDFs of age and
extinction. In cases where one $Z$ is highly preferred, the lack of a
finer grid leads to an underestimation of the statistical
uncertainties in the SC parameters. Experiments with $Z$ interpolated
spectra alleviate these problems to some extent, but the ultimate
solution is to have a finer sampling in both $Z$ and $t$.

\begin{figure}
\includegraphics[bb= 30 160 580 710,width=0.45\textwidth]{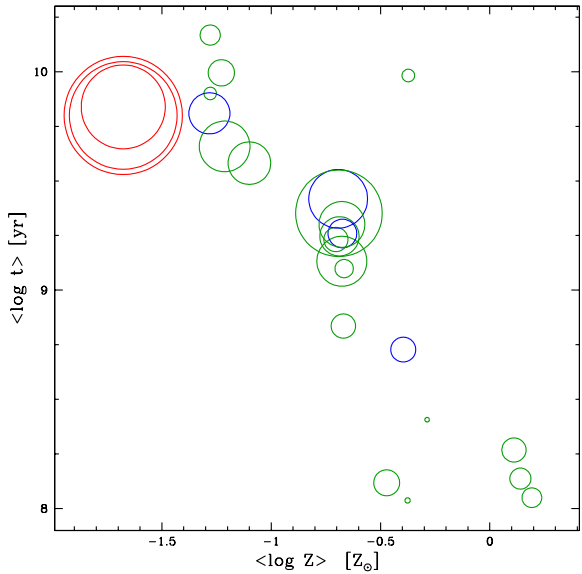}
\caption{Multi-SSP age dispersion plotted in the $t$-$Z$ plane, where
$\langle \log t \rangle$ and $\langle \log Z/Z_\odot \rangle$ given by
our Bayesian estimates.  The size of the circles scales with
$\sigma_m$ (the age dispersion in the multi-SSP fits), which ranges
from 0 to 0.9 dex. Red circles correspond to cases where multi-SSP
fits are much better than single SSP ones ($\delta_m > 20\%$), while
green is used when $\delta_m < 10\%$ and blue correspond to cases in
between.  Large $\sigma_m$, like those seen for leftmost sources (M15,
M79 and NGC 2010), is indicative of a component not adequately
represented in the models, like blue HB stars.  NGC 1818 is not
plotted because it is younger than the lower age in the V00s models.}
\label{fig:HBtest}
\end{figure}

A curious effect found in some of the multi-SSP fits was a split into
a light dominating old population ($\sim 80\%$) and weaker ($\sim
20\%$) but significant population of much younger age (often the
lowest age in the grid), as if the SSP model spectra lacked old blue
stars. In principle, these multi-component {\sc starlight} fits should
not be relevant for SCs, but we find that they are able to capture
symptoms of inadequate single SSP fits in a quantitative
manner. Clusters with suspicious combinations of different ages may be
identified by large values of $\sigma_m$ (ie., large dispersions in
$\log t$). Four clusters have $\sigma_m > 0.6$: M15, M79, NGC 2203,
and NGC 2210 (Figs.\ \ref{fig:ExampleFit2} and
\ref{fig:All3Contours_NGC2210}).  M15 and M79 are very similar to NGC
2210. In fact, they all have identical best fit $t$ and $Z$, as well
as the same $\log T_m$, similar $f_m$, and $\delta_m$ values which
indicate that the multi-SSP fit is substantially better than a single
SSP fit ($\delta_m = 44\%$ for M15, $97\%$ for M79, and $27\%$ for NGC
2210).  NGC 2203, on the other hand, is only similar to NGC 2210 in
terms of $\sigma_m$, and its multi-SSP fit is not substantially better
than single SSP one ($\delta_m = 8\%$).  Thus, the ``old plus young''
anomaly is only convincingly detected in M15, M79 and NGC 2210.

Following K08, we suggest this result is due to incomplete
modelling of the horizontal branch (HB). CMD work on M15 (Battistini
et al.\ 1985; Buonanno et al.\ 1983), M79 (Kravtsov et al.\ 1997) and
NGC2210 (Brocato et al. 1996) have shown that all these clusters have
blue HBs, which supports our interpretation of the spectral synthesis
results.  Alternatively, blue stragglers would have a similar effect.
Combinations of UV and optical photometry detect the presence of these
stars in M15 (Guhathakurta et al.\ 1993; Ferraro \& Paresce 1993) and
M79 (Lanzoni et al.\ 2007). The situation is less clear in NGC 2210.
Based on the similarity of NGC 2210 HST CMD to NGC 2257, Johnson et
al.\ (1999) suggest that the cluster have both a blue HB and blue
stragglers, while Brocato et al (1996) have argued (on the basis of a
ground based CMD) that the stars in the blue straggler region are
likely field contaminants in NGC 2210.

In broad terms, one expects bluer HB at lower $Z$ (Harris 1996;
Greggio \& Renzini 1990), and hence the age-spread noticed in the
multi-SSP fits should increase for decreasing $Z$.  Fig.\
\ref{fig:HBtest} presents tentative evidence of this relation.  SCs
with large $\sigma_m$ and $\delta_m$ tend to be those with smallest
$Z$. Because $Z$ and $t$ are strongly anti-correlated in this sample,
the highest $\sigma_m$ clusters are also among the oldest ones. Given
that $Z$ is not the single factor determining HB morphology, not to
mention other possibilities (like blue stragglers and stochastic
effects) and uncertainties involved, we shall not elaborate further
on this issue.

\begin{figure}
\includegraphics[bb= 60 170 300 590,width=0.45\textwidth]{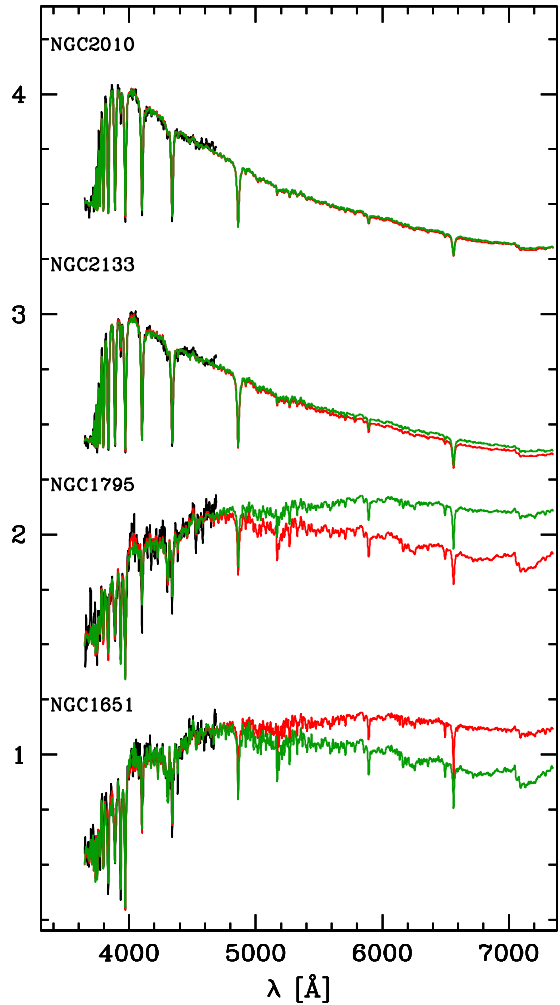}
\caption{Extension of statistically indistinguishable fits to the
3650--4600 \AA\ data to the red limit of the models for 4 SCs.  For
each cluster, the red and green lines are fits with different
metallicities ($\log Z/Z_\odot = -1.3$ and $-0.7$ for NGC 1651 and NGC
1795, and $\log Z/Z_\odot = 0$ and $+0.2$ for NGC 2133 and NGC
2010). The observed spectra are plotted in black.}
\label{fig:WIDEtest}
\end{figure}

Moving away from caveats in the stellar population issues, we must
acknowledge at least one limitation of our results: Wavelength
coverage. The spectra used here cover a relatively small region in the
blue--near-UV. Not only it misses some strong $t$ and $Z$ sensitive
absorption features, like H$\beta$ and the MgI lines, but the lack of
a wider baseline prevents a firmer handle on $A_V$. No doubt a wider
coverage would help better constrain SC properties, but how much?
Could more data help, for instance, lifting the ambiguities in $Z$
found for several SCs?  

A qualitative answer to this question is given in
Fig.~\ref{fig:WIDEtest}, where, for each of four SCs, we plot the two
best SSP fits with different $Z$, extending it from the 3650--4600
\AA\ range covered by the data to the red limit of the V00s models.
For the two clusters at the bottom of the plot, NGC 1651 and NGC 1795,
for which $\log Z/Z_\odot = -1.3$ and $-0.7$ provide statistically
indistinguishable fits, the extrapolated model spectra differ a lot
both in continuum shape and in the depth of absorption features to the
red of 4600 \AA. In these cases, data in the red would clearly help
distinguishing the fits.  For NGC 2010 and NGC 2133, on the other
hand, spectra in the red would not be as useful, as the extrapolated
models are essentially indistinguishable. A wider baseline would be
needed to pin down the best solution in these younger systems.

\section*{Acknowledgments}

This work has been funded with support from the Spanish Ministerio de
Educaci\'on y Ciencia through the grants AYA2007-64712, and
co-financed with FEDER funds.  We are grateful to Alexandre
Vazdekis for making publicly available his models in advance of
publication, James Rose who kindly sent us the stellar cluster
spectra.  We also thank Jo\~ao Francisco dos Santos, Miguel Cervi\~no,
Jes\'us Ma\'\i z-Apell\'aniz, Claus Leitherer, David Valls-Gabaud and
Eduardo Bica for discussions and advice.  We also thank support from a
joint CNPq-CSIC bilateral collaboration grant.  RGD thanks too Cid's
family for their support and hospitality along the years.

\appendix

\section{Interpolations in metallicity}
\label{app:Appendix_ZIwork}

In the absence of a fine $Z$-grid, one may resort to
interpolations. Because of stellar evolution effects, interpolating in
$Z$ at fixed $t$ is not strictly valid, but one expects it to work at
least approximately.  This appendix describes results of experiments
with $Z$-interpolated models. The original $N_Z = 6$ grid was
interpolated to $N_Z = 39$ with $\log Z/Z_\odot$ between $-1.7$ and
$+0.2$ in 0.05 dex steps.  The spectral interpolations were carried
out in log-log space, with $\log F_\lambda(t;Z) = a_\lambda(t) \log
Z/Z_\odot + b_\lambda(t)$.

\begin{figure}
\includegraphics[bb= 30 180 550 710,width=0.45\textwidth]{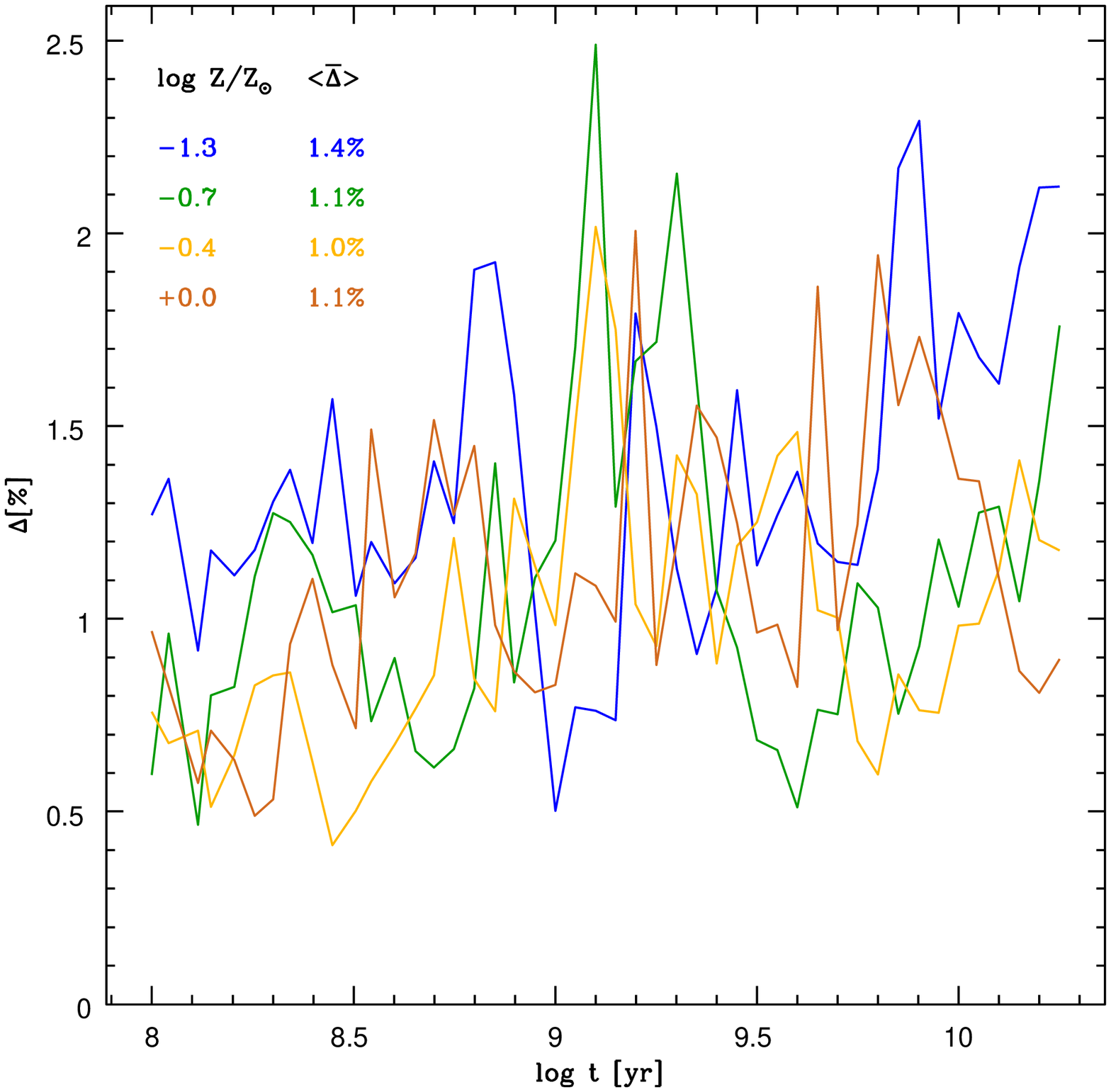} 
\caption{Mean percentage spectral residual ($\overline{\Delta}$) as a
function of age for $Z$-interpolated models as compared to the
original ones for $\log Z/Z_\odot = -1.3$, $-0.7$, $-0.4$ and $0.0$.
Average values of $\overline{\Delta}$ for all ages are listed in the
inset table.  The message from this plot is that interpolating SSP
spectra in $Z$ introduces errors which are of the same order as the
spectral residuals obtained in fits of actual SC data.}
\label{fig:ZI_spectral_residuals}
\end{figure}

To test how well $Z$-interpolation works we have compared the original
$\log Z/Z_\odot = -1.3$, $-0.7$, $-0.4$ and $0.0$ spectra with those
obtained through interpolation of models with adjacent $Z$-values (for
instance, using the $-1.3$ and $-0.4$ models to interpolate to
$-0.7$). The results are shown in Fig.\
\ref{fig:ZI_spectral_residuals}, where we plot the mean percentage
spectral deviation $\overline{\Delta}$ against $t$ for each of these
four intermediate $Z$'s. The residuals are computed after
re-normalizing the spectra at 4020 \AA, eliminating differences in the
absolute flux scale.  The plot shows that interpolation leads to
typical errors of 1\% in terms of spectral residuals, but that in many
cases $\overline{\Delta}$ approaches or even exceeds 2\%.  These
residuals are of the same order of the $\overline{\Delta}$ values
obtained in fits of actual SCs (Table \ref{tab:Results_BVM00s})!
Naturally, in these tests the interpolations are based on models
farther apart in $Z$ than when using the full grid. Nevertheless, we
take these results as a signal that an interpolated grid may not lead
to as much improvement as one would hope to achieve.

\begin{figure}
\includegraphics[bb= 50 180 280 710,width=0.45\textwidth]{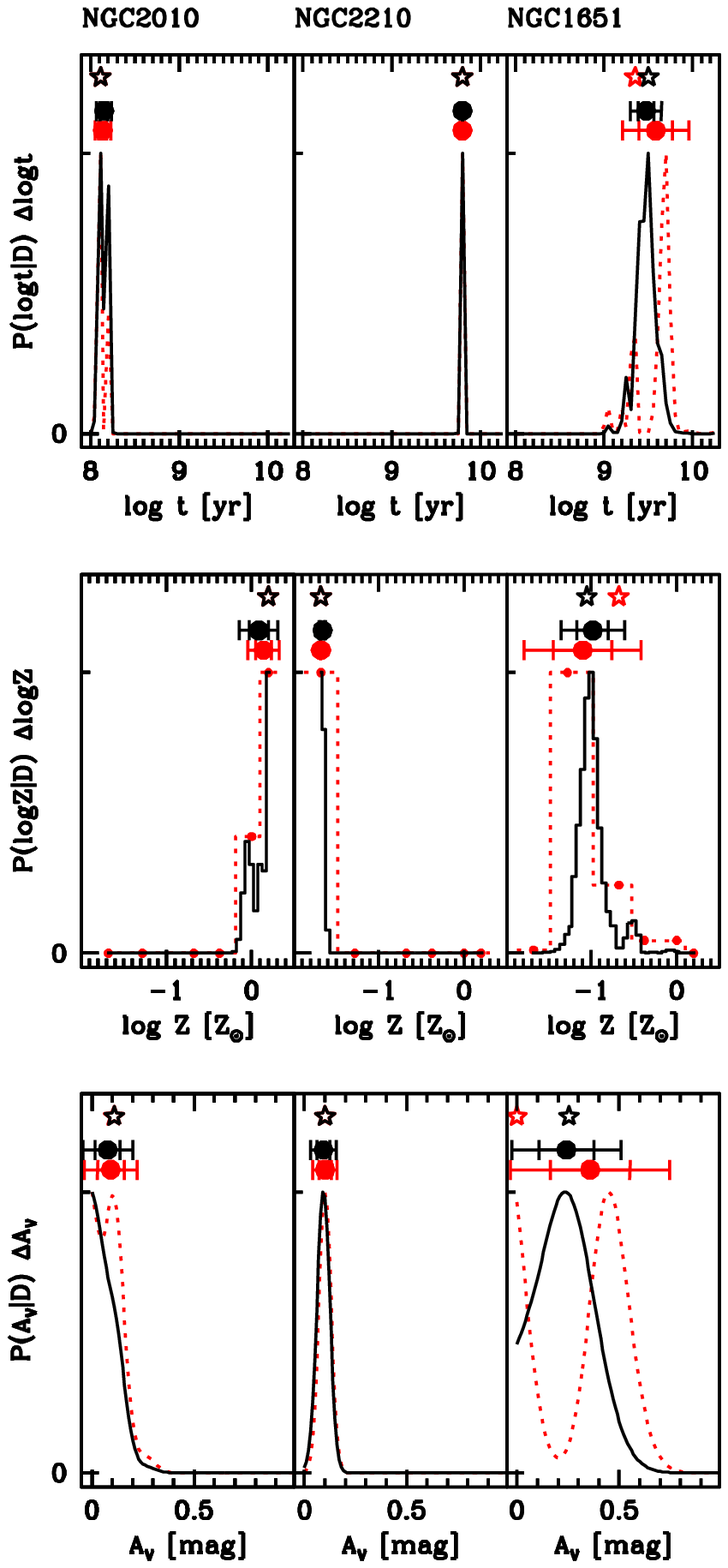}
\caption{Comparison of the $t$, $Z$ and $A_V$ PDFs obtained with the
original (dotted red lines) and $Z$-interpolated grids (solid black),
for three SCs in our sample. As in Fig.\ \ref{fig:PDFs_n2010}, a star
marks the best-fit value, and the filled circle marks bayesian
estimate. Error bars are also shown.}
\label{fig:ZI_PDFs}
\end{figure}

We now check how a finer $Z$-grid changes our parameter estimates.
Fig.\ \ref{fig:ZI_PDFs} shows the marginalized PDFs of $t$, $Z$ and
$A_V$ for NGC 2010, NGC 2210 and NGC 1651. The plot compares the PDFs
derived with the original $N_Z = 6$ Z-grid (dotted red lines) with
those derived with the $N_Z = 39$ interpolated models (solid black).
For NGC 2010 and NGC 2210 the $Z$-interpolated PDFs do not differ
significantly from the original ones. The clearest changes are in the
PDF($A_V$) for NGC 2010, which becomes smoother (single-peaked), and
in the PDF($Z$) for NGC 2210, which rises steeply towards $\log
Z/Z_\odot = -1.7$ (confirming that the lowest $Z$ models are indeed
the ones which better fit the data for this SC). More significant
changes occur in the case of NGC 1651. With the $Z$-interpolated
models, PDF($t$) has a pronounced peak in between the two peaks
obtained with the original grid. Similarly, the PDF($Z$) now peaks in
between the two original $Z$ values with highest probability. The more
dramatic change is in PDF($A_V$), whose original two peaks merge into
a single one at an intermediate value of $A_V$. For all three
parameters the $Z$-interpolated models produce more focused PDFs, and
thus smaller formal uncertainties (see error bars in the NGC 1651
panels of Fig.\ \ref{fig:ZI_PDFs}). Notwithstanding these apparently
large differences, the bayesian estimates of $t$, $Z$ and $A_V$
derived with the coarser original $Z$-grid are consistent with those
derived with the better sampled interpolated $Z$-grid. They are also
more conservative, in the sense of having larger error bars.

Considering the sample as a whole, we detect no bias between
parameters estimated with the original and $Z$-interpolated grids.
The rms difference in the bayesian estimates of $\log t$ and $\log
Z/Z_\odot$ are just 0.08 and 0.12 dex respectively, while for $A_V$
the rms difference is 0.08 mag. $Z$-interpolated grids produce better
spectral fits, but only marginally so. In terms of the
$\overline{\Delta}$ figure of merit, the $Z$-interpolated fits yield a
sample-average value of $\overline{\Delta} = 2.39\%$, compared to
2.44\% obtained with the original grid.

Regarding formal (PDF-based) parameter uncertainties, our experiments
show that when an uncertainty derived with the original grid is
``large'' ($\ga 0.05$ dex in $\log t$, $\ga 0.15$ dex in $\log
Z/Z_\odot$, and $\ga 0.1$ mag in $A_V$), it tends to be smaller with
the $Z$-interpolated grid, and vice versa.  This behavior is
consistent with qualitative expectations.  In cases where a PDF is
unresolved due to a coarse grid, leading to a formal error of zero,
interpolation resolves the PDF, producing a measurable second moment
(standard deviation $> 0$).  For instance, for NGC 2210 we find
$\sigma(\log Z) = 0$ with the original grid (all probability is in the
first $Z$-bin), and 0.02 dex with the interpolated one. Conversely,
when, say, the PDF of $Z$ is spread over two or more adjacent original
$Z$-bins, a finer grid may produce a more focused (narrower) PDF, as
in the case of NGC 1651. In any case, these are relatively small
differences, such that the sample-average uncertainties derived with
both methods are virtually indistinguishable.

All in all, we see no major gain in resorting to $Z$-interpolated
models. Further considering the uncertainties introduced by such
interpolations (Fig.\ \ref{fig:ZI_spectral_residuals}), we think it is
safer to wait for properly computed, finer $Z$-grids to be released by
than to interpolate in $Z$.


\section{Effects of data quality}
\label{app:Appendix_SNwork}

The {\sc starlight} code, used throughout this paper, has been
extensively used and tested in the context of mixed stellar
populations (galaxies), but this is the first time it is used to fit
SCs.  This appendix presents simulations designed to evaluate the
reliability of SC properties derived with this code under different
levels of noise.

To test the effects of data quality, we have performed simulations in
which white gaussian noise is added to a model SSP spectrum of known
$t$, $Z$ and $A_V$. These perturbed spectra are processed exactly as
the observed ones, and input and output parameters are then compared.
The input models were built using the V00s SSPs which best fit the 27
actual SCs in our sample, whose properties are listed in columns 2--4
of Table \ref{tab:Results_BVM00s}. Their corresponding spectra were
taken directly from {\sc starlight}'s output\footnote{Only version 5
of the code outputs the best fit SSP spectrum.}, and thus mimic the
actual data in terms of wavelength coverage and resolution. Five
realizations of a noise spectrum ($n_\lambda$) were generated, and
five others were derived from these by inverting the sign of
$n_\lambda$. These 10 perturbation spectra were scaled to emulate
$S/N$ ratios between 10 and 95.

\begin{figure}
\includegraphics[bb= 40 210 535 600,width=0.45\textwidth]{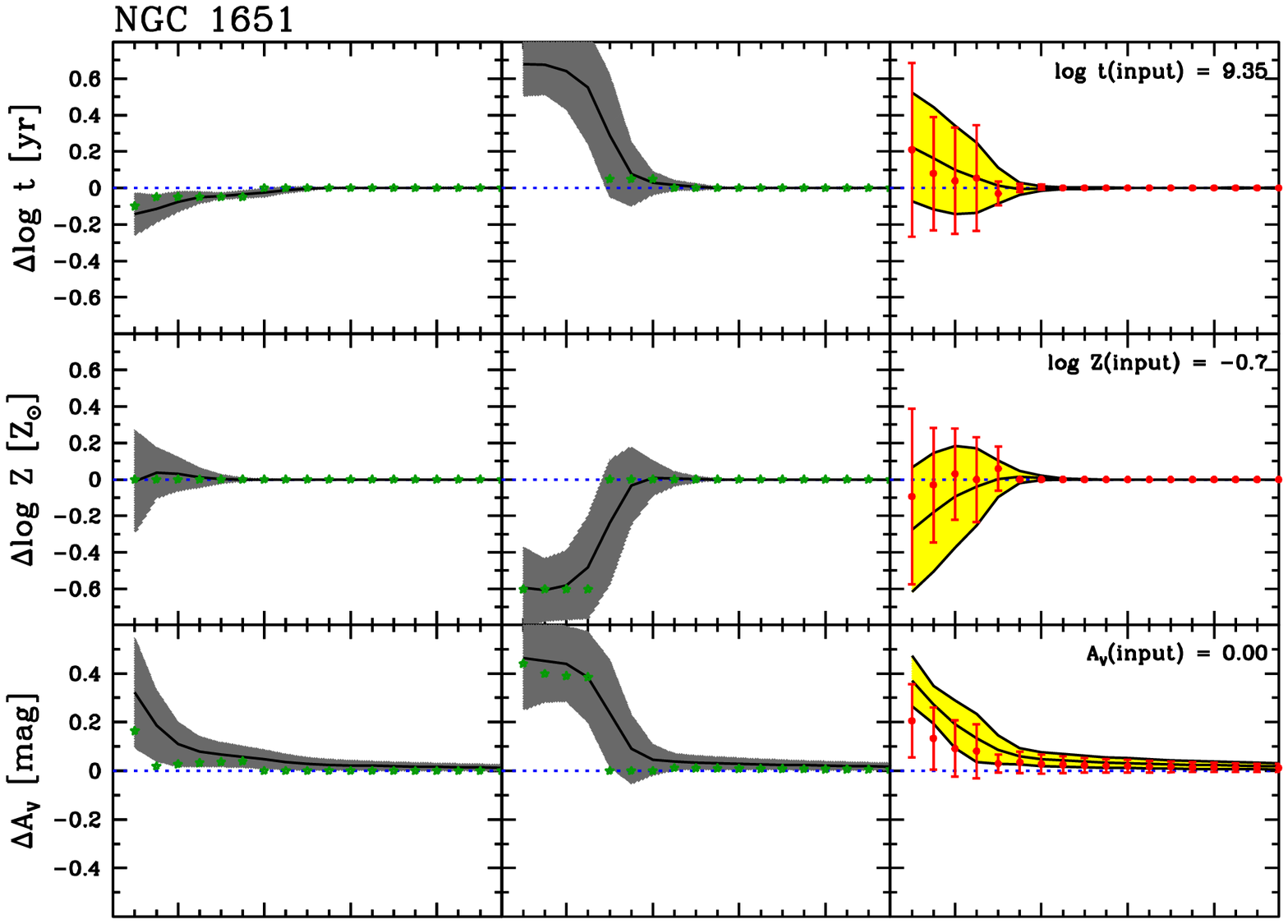}
\includegraphics[bb= 40 210 535 600,width=0.45\textwidth]{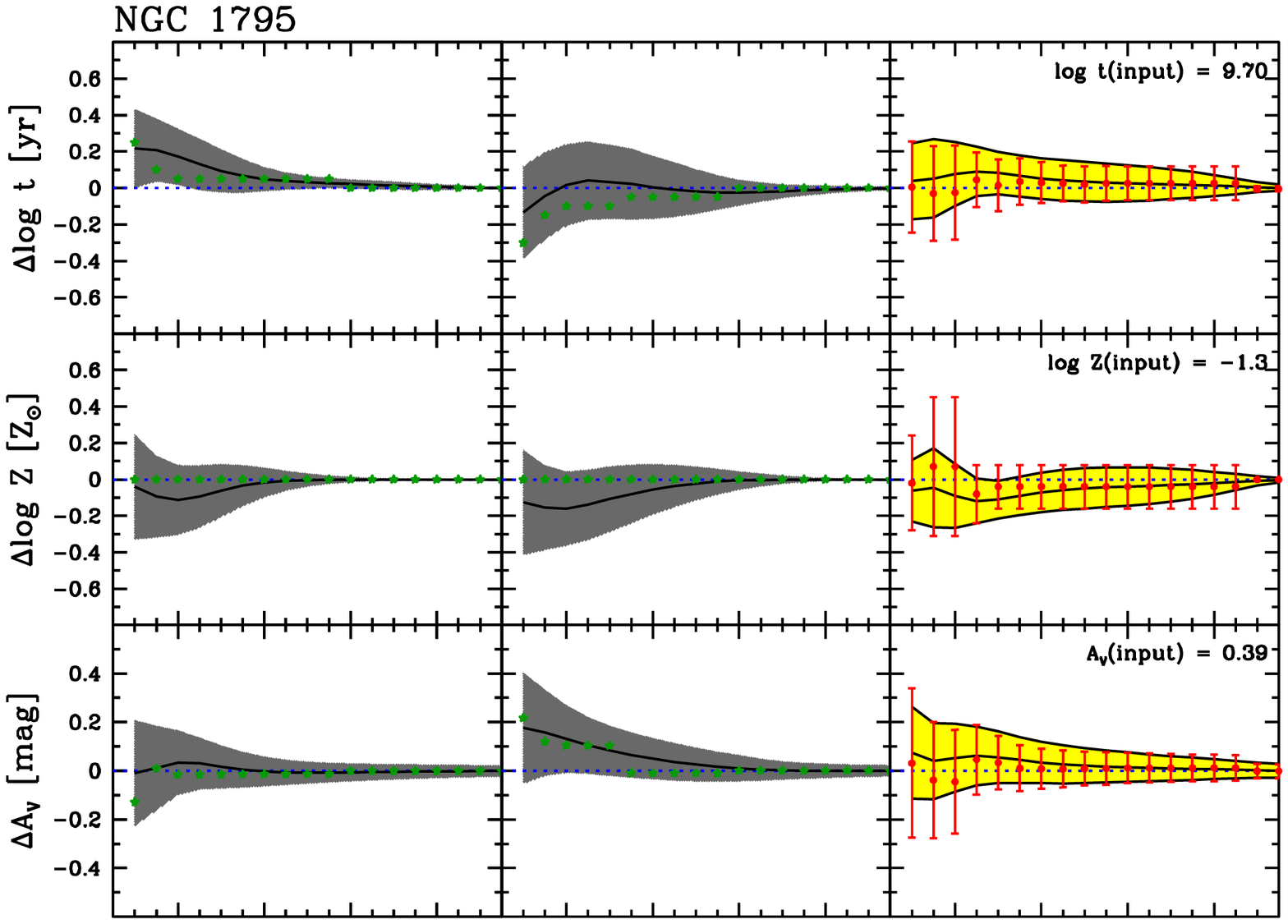}
\includegraphics[bb= 40 180 535 600,width=0.45\textwidth]{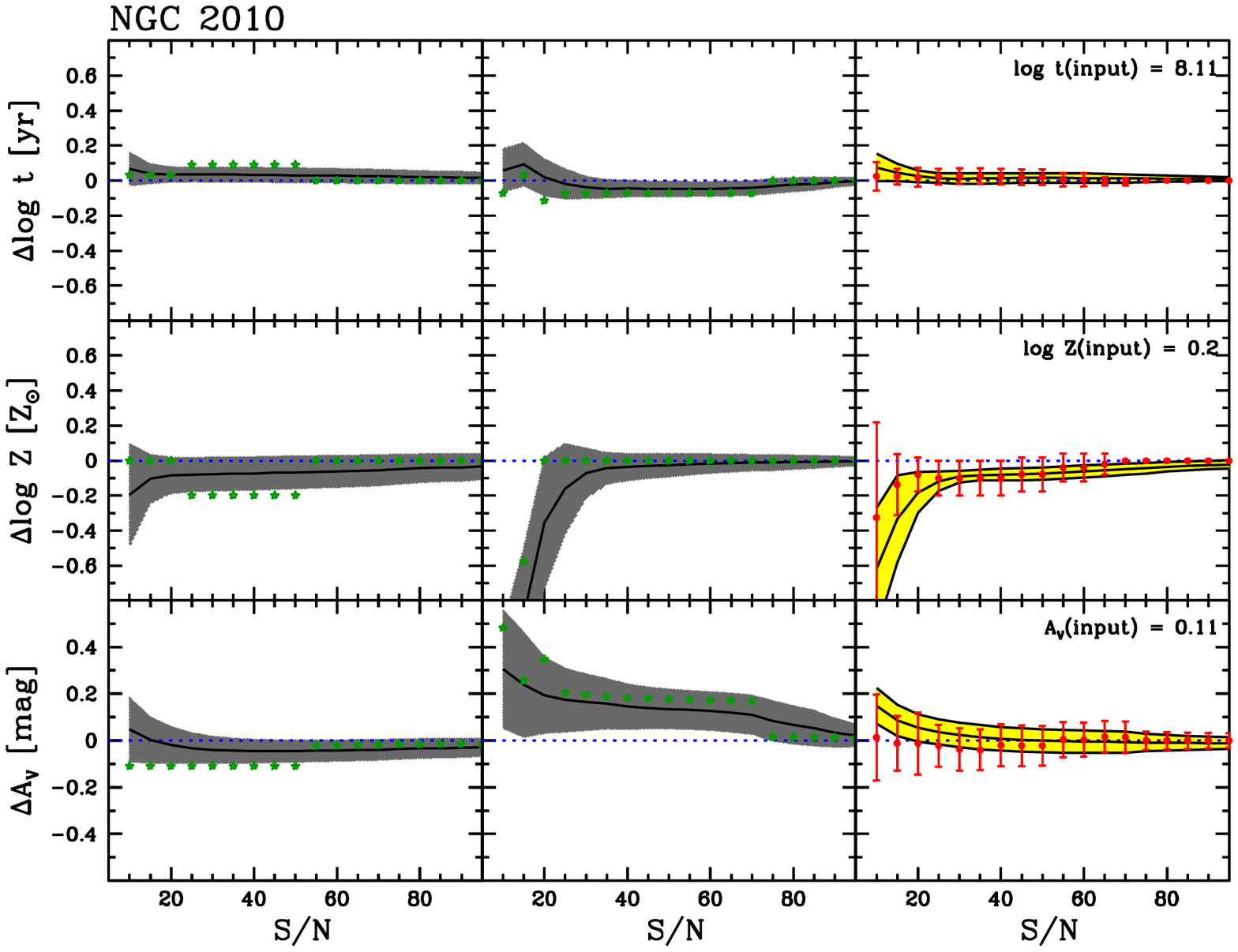}
\caption{Illustration of the effects of noise upon derived SC
properties for simulations based on the $t$, $Z$ and $A_V$ values
obtained for NGC 1651 (top panels), NGC 1795 (middle) and NGC 2010
(bottom).  The 1st and 2nd columns of plots show results of the
analysis of two individual perturbed spectra. Error bars and the
shaded area show the PDF based $\log t \pm \sigma(\log t)$, and
similarly for $\log Z$ and $A_V$, while the (green) star marks the
best-fit value.  The last column shows the statistics of the
simulations. In this case the shaded area shows the mean and $\pm$ one
sigma range of the PDF based values of $\overline{\log t}$,
$\overline{\log Z}$ and $\overline{A_V}$. The points with error bars (in
red) show the mean and standard deviation of the best-fit values among
10 realizations of the noise for each $S/N$.}
\label{fig:SN_simulations}
\end{figure}

Fig.\ \ref{fig:SN_simulations} illustrates the results for three test
SCs, with parameters based on those found for NGC 1651, NGC 1795 and
NGC 2010.  All panels show output minus input parameters against the
$S/N$.  The left and middle columns show, for each $S/N$, the results
of an analysis based on an {\em individual} spectrum.  The examples
picked for these panels differ only in the sign of the perturbation,
such that if in the left column the spectrum feed into {\sc starlight}
is $F_\lambda + n_\lambda$, in the middle one the analysis is carried
out with $F_\lambda - n_\lambda$.  Also, $n_\lambda$ has the same
shape for all $S/N$ values, such that the only thing that changes
along the $x$-axis is the amplitude of the noise. This cosmetic choice
is done only for presentation purposes, as it produces more
illustrative plots than obtained picking perturbed spectra at
random. Finally, the third column of plots shows statistical results
considering all 10 spectra available at each $S/N$.

Consider first the left and middle panels.  Stars show the difference
between best fit and input values of $\log t$, $\log Z$ and $A_V$ as a
function of $S/N$. The solid line shows the differences between the
bayesian estimates and the input values, while the grey shaded areas
show the corresponding $\pm$ one sigma formal (PDF-based)
uncertainty. Besides the expected convergence as $S/N$ increases,
these plots illustrate covariances between age, metallicity and
extinction. When the $Z$-estimate does not change, underestimated
$t$'s are compensated by overestimated $A_V$'s, and vice-versa, as in
the left panels for NGC 1651 and in the left and middle panels for NGC
1795. In the middle panel for NGC 1651, one sees that both $t$ and
$A_V$ are overestimated at low $S/N$ due to the fact that fits with
the $\log Z/Z_\odot = -1.3$ grid point are preferred over the $\log
Z/Z_\odot = -0.7$ original value. In this particular example, only for
$S/N \ge 30$ the input parameters are well recovered. In the case of
NGC 2010, the left panels show that solutions oscillate between $Z =
1.5 Z_\odot$ (the input value) and $Z_\odot$, and all SC parameters
vary in concert. This metallicity confusion occurs for $S/N$ as large
as 50, which is not surprising, given the high degree of spectral
similarity between solar and over-solar V00s models in this age range
(see Fig.\ \ref{fig:AVrmsAndFits_n2010}). In the example examined in
the middle panels, $Z$ comes out badly underestimated for $S/N \le
15$. Above that $Z$ is well recovered, but minor differences in $A_V$
and $t$ can persist to $S/N$ as high as 70.

These examples illustrate what may happen in the analysis of {\em
individual} spectra observed under different regimes of $S/N$.  The
general message spelt by these simulations is that, when dealing with
one single object and using only the kind of observational data used
in this paper, one should be careful about biases in parameter
estimates obtained through spectral synthesis if $S/N \la 30$. Beyond
this $S/N$, output minus input deviations are too small to worry
about.

The right panels in Fig.\ \ref{fig:SN_simulations} show the results of
a statistical analysis of all perturbed spectra for each test-SC. The
yellow shaded ranges marks the dispersion in the bayesian estimates of
$\overline{\log t}$, $\overline{\log Z}$, and $\overline{A_V}$ among
the 10 perturbed versions of a same spectrum.  Similarly, the red
error bars show the dispersion in the best fit estimates as a function
of $S/N$.  The sometimes large deviations observed in the left and
middle panels average out in these statistical results. Biases in the
estimates are still present at low $S/N$, but almost always within one
sigma of output $=$ input. The most noticeable bias in the right
panels of Fig.\ \ref{fig:SN_simulations} occurs for $A_V$ in NGC 1651,
and only for low $S/N$. Since $A_V < 0$ was forbidden in the fits, and
the input value of $A_V$ is exactly zero in this case, one can only
err upwards, so this bias can be understood as an edge effect.

\begin{figure}
\includegraphics[bb= 40 180 305 700,width=0.45\textwidth]{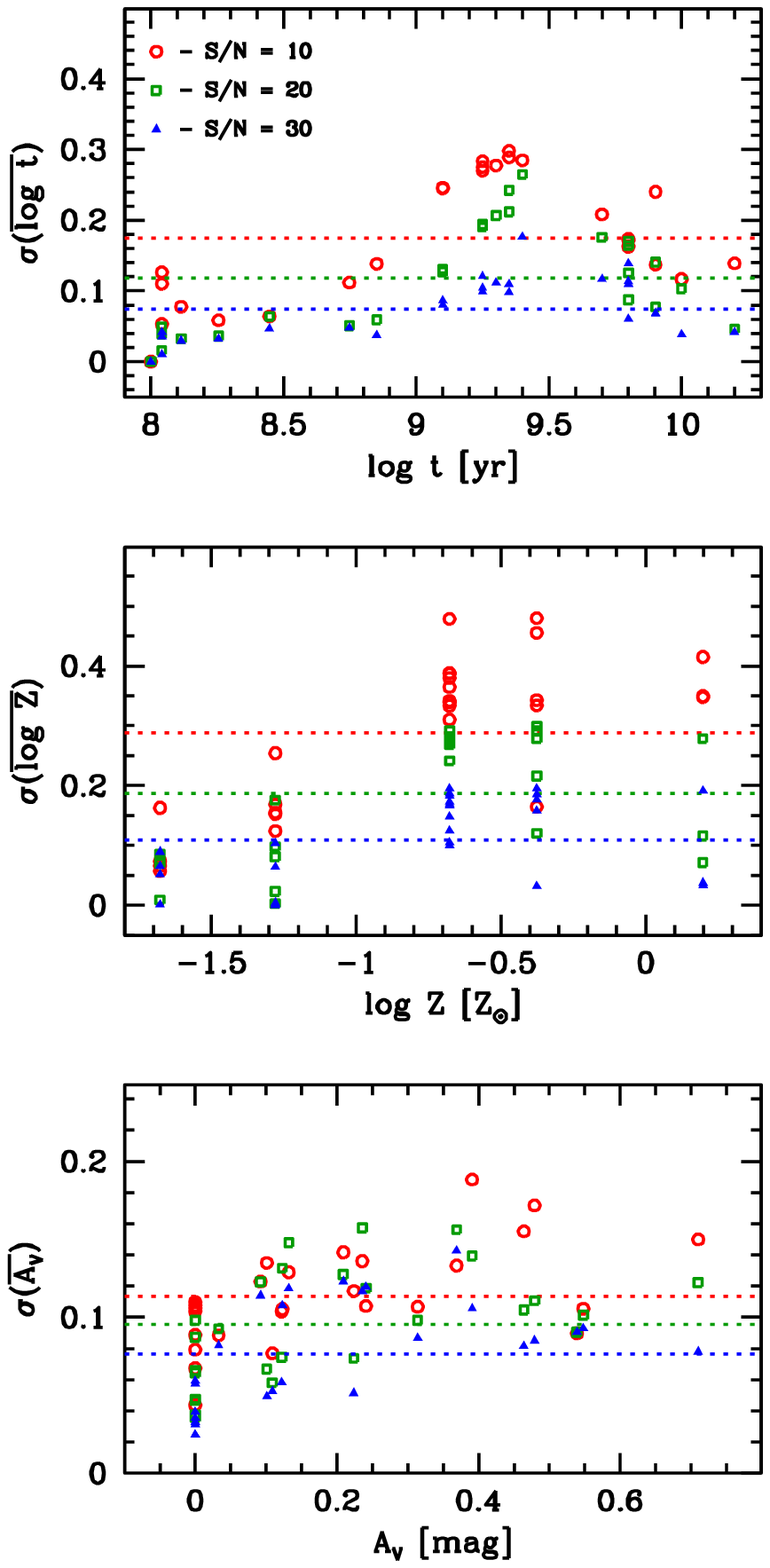}
\caption{Statistical uncertainties on $\overline{\log t}$ (top panel),
$\overline{\log Z}$ (middle) and $\overline{A_V}$ (bottom) for
simulations with $S/N = 10$ (red open circles), 20 (green open
squares) and 30 (blue solid triangles). Dotted horizontal lines mark
the mean value over all 27 simulated clusters.  Each point
corresponds to results from 10 Monte Carlo realizations of the
noise. The 27 points (for each $S/N$ level) correspond to test
clusters whose parameters are identical to those in columns 2--4 of
Table \ref{tab:Results_BVM00s}.}
\label{fig:SN_Stats}
\end{figure}

Fig.\ \ref{fig:SN_Stats} summarizes the statistics of the
noise-induced variations in bayesian parameter estimates for all 27
test SCs. The top panels, show the standard deviation (considering the
10 perturbations) of the $\overline{\log t}$ estimate against the
input value of $\log t$ for three values of $S/N$: 10 (red open
circles), 20 (green open squares) and 30 (blue filled
triangles). Dotted lines mark the average $\sigma(\overline{\log t})$
for these 3 values of $S/N$.  The plot shows that the typical
uncertainties in $\overline{\log t}$ due to noise are of $< 0.2$ dex
for $S/N > 10$. For SCs in the 9.0 to 9.5 $\log t$ range the
uncertainty reaches 0.3 dex for $S/N = 10$, which is acceptable
considering such low quality data. The middle and bottom panels show
the results for $\log Z/Z_\odot$ and $A_V$.  For the metallicity,
noise at a $S/N = 10$ level introduces up to 0.5 dex uncertainty, but
the overall average value is 0.3 dex. As for the ages, the uncertainty
in $Z$ peaks at intermediate values.  A plausible explanation for this
is that in the middle of a ($t$ or $Z$) grid one can look to either
side and find reasonably similar spectra, whereas closer to the edges
there is not so much option, thus reducing the ranges in $t$ and $Z$
where statistically acceptable fits can be found.  For $A_V$ we find
up to 0.2 mag uncertainty, with a typical value of 0.1 mag.

The overall conclusion of these experiments is that, for the kind of
data used in this paper (medium resolution spectra in the blue range),
$S/N \ga 30$ is required to accurate SC parameter estimates from
spectral synthesis of an individual object. When large samples of
objects are analyzed, one can trust statistical results obtained with
lower quality data.

\begin{figure}
\includegraphics[bb= 40 160 535 710,width=0.45\textwidth]{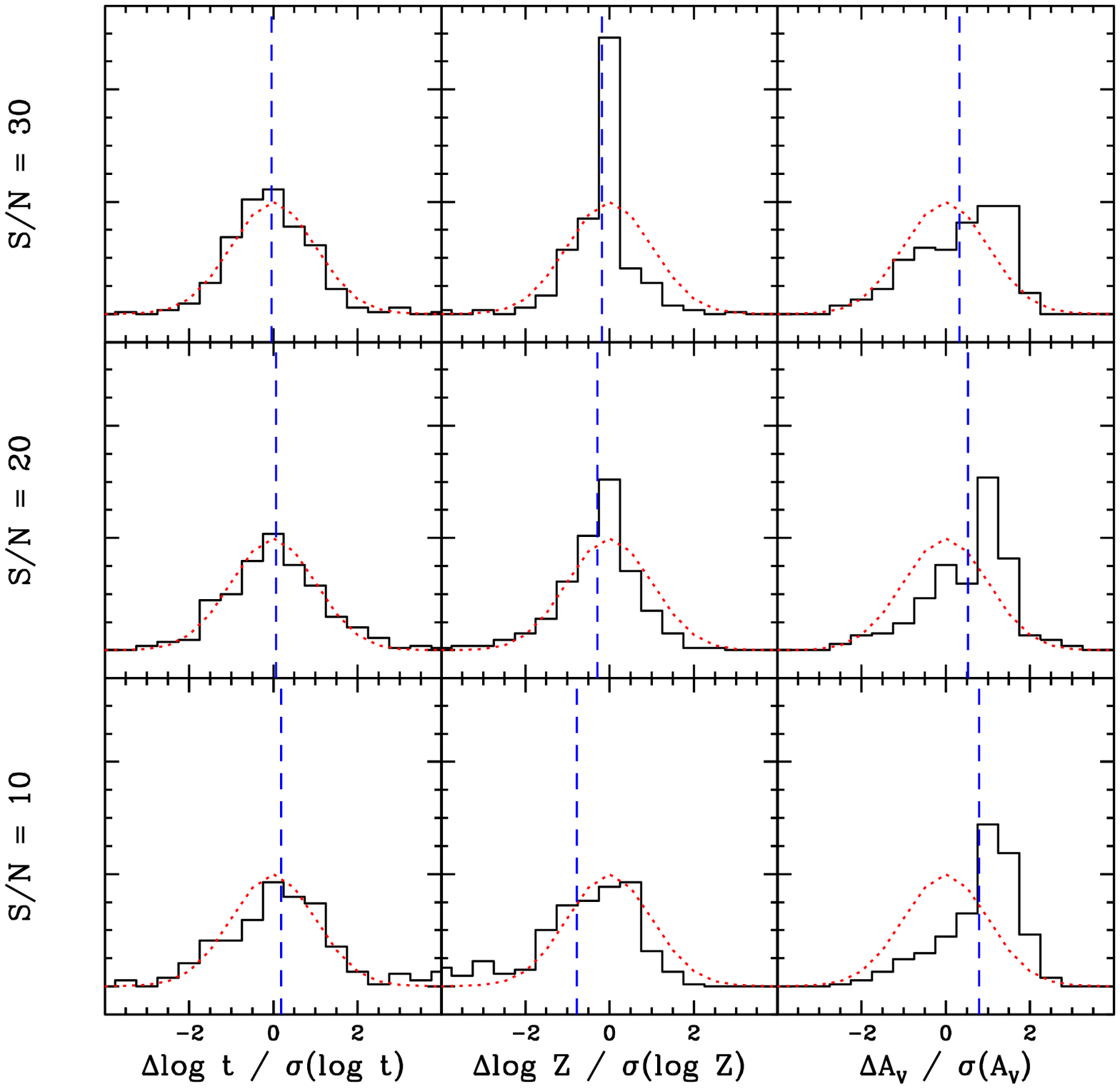}
\caption{Distribution of the difference between the output bayesian
estimate and the input parameter in units of the formal uncertainty
for $\log t$ (left), $\log Z$ middle, and $A_V$ (right).  Vertical
dashed lines mark the average value of the normalized
difference. Dotted lines (in red) correspond to gaussians of unit
standard deviation and normalized to the same area as the simulated
data. The histograms include results for 270 runs, corresponding to 10
perturbations of the 27 test-SCs. Botttom, middle and top panels
correspond to simulations with $S/N = 10$, 20 and 30, respectively.}
\label{fig:SN_ReducedDifs}
\end{figure}

We emphasize that our conclusion that $S/N \ga 30$ is required to
produce accurate parameter estimates is {\em not} equivalent to saying
the method fails at lower $S/N$. Our simulations show that, as
expected, output minus input differences in SC properties tend to
increase as $S/N$ decreases, {\em but so do the associated error
bars}. The reliability of a parameter estimation method is better
measured not in terms of absolute differences, but in terms of
differences in units of the associated uncertainty, e.g.,
$[\overline{\log t} - \log t(\rm input)] / \sigma(\log t)$. Evaluated
in this way, spectral synthesis is capable of producing good results
for any of the $S/N$ investigated. This is illustrated in Fig.\
\ref{fig:SN_ReducedDifs}, which shows histograms of output minus input
parameters in units of the corresponding formal uncertainty. All 10
perturbed versions of all 27 SCs are included in these
histograms. Bottom, middle and top panels show results for $S/N = 10$,
20 and 30, respectively. Dotted lines draw gaussians of unit $\sigma$
and normalized to the same area. The histograms follow quite well the
the gaussian expected in the ideal case, with the noticeable exception
of $A_V$, which is systematically overestimated as a result of the
already mentioned $A_V \ge 0$ edge effect. For $S/N = 10$ this bias is
close to 1 sigma, as can be seen by the location of the vertical
dashed line, which marks the average value of $[\overline{A_V} -
A_V(\rm input)] / \sigma({A_V})$.  For $S/N = 20$ this difference
reduced to 0.5 sigma on average, while for better data the
distribution of $[\overline{A_V} - A_V(\rm input)] / \sigma({A_V})$
becomes progressively less skewed.  Another effect which can be
noticed in this plot is that for high $S/N$ the distribution of
$[\overline{\log Z} - \log Z(\rm input)] / \sigma(\log Z)$ becomes
narrower than expected. This happens because as the noise decreases
the PDF of $Z$ becomes increasingly focused and centered on the
correct grid-value of $Z$, and hence $\overline{\log Z} = \log Z(\rm
input)$.

\end{document}